\documentclass{aa}  

\usepackage{graphicx}
\usepackage{txfonts}
\usepackage{booktabs} % put this in the preamble
\usepackage[normalem]{ulem}
\usepackage{subfig}
\usepackage{xcolor}

\usepackage{hyperref}
\begin{document}

   \title{Testing modified gravity with 3x2pt analyses in galaxy mocks}

   \author{M. Alemany-Gotor
          \inst{1,2}\thanks{E-mail: alemany@ice.csic.es}
          \and
          C. Viglione\inst{1,2}
          \and
          P. Fosalba\inst{1,2}
          \and
          I. Tutusaus\inst{1,2,3}
          }

   \institute{Institute of Space Sciences (ICE, CSIC), Campus UAB, Carrer de Can Magrans, s/n, 08193 Barcelona, Spain
         \and
             Institut d'Estudis Espacials de Catalunya (IEEC), Edifici RDIT, Campus UPC, 08860 Castelldefels (Barcelona), Spain      
        \and
    Institut de Recherche en Astrophysique et Plan\'etologie (IRAP), Universit\'e de Toulouse, CNRS, UPS, CNES, 14 Av. Edouard Belin, 31400 Toulouse, France}

   \date{Received X, X; accepted X}

% \abstract{}{}{}{}{} 
% 5 {} token are mandatory
 
  \abstract
  {Stage-IV surveys will enable unprecedented tests of gravity on cosmological scales. However, assuming General Relativity in the analysis of large-scale structure could introduce systematic biases if gravity deviates from GR at these scales. Modified gravity theories, such as the Hu-Sawicki formulation of $f(R)$ gravity, offer an alternative explanation for cosmic acceleration without invoking a cosmological constant, while remaining consistent with Solar System tests through screening mechanisms. In this work, we quantify the cosmological parameter biases that arise when using a combination of galaxy clustering and weak-lensing data-vectors, the so-called 3x2pt analysis, from an $f(R)$ galaxy mock under the incorrect assumption of GR, using for the first time high-fidelity full-sky galaxy mock catalogues. We employ a pair of twin simulations: one with GR and one with Hu--Sawicki $f(R)$ gravity with $|f_{R0}| = 10^{-5}$. The mocks are built using an HOD method to populate the dark matter haloes with galaxies, calibrated against SDSS observations at low redshift. Using conservative scale cuts to minimise modelling uncertainties, we perform 3x2pt analyses and infer cosmological parameters through nested sampling, validating our pipeline with the GR mock. Our results show that when analysing the $f(R)$ galaxy mock assuming GR, the recovered cosmological parameters are very significantly biased, even when considering conservative scale cuts: the Figure of Bias reaches $\sim12\sigma$ for both $\{\Omega_{\rm m}, \sigma_8\}$ and $S_8$. These biases persist even when marginalising over the galaxy bias and baryonic feedback, demonstrating that nuisance parameters cannot absorb the effects of modified gravity. We conclude that incorrectly assuming GR in a universe governed by $f(R)$ gravity leads to severe and detectable biases in cosmological inference for Stage-IV surveys.}


   \maketitle
%
%-------------------------------------------------------------------

\section{Introduction}
\label{sec:intro}

Cosmology has entered an era of precision driven by stage-IV surveys. The unprecedented amount and quality of the data that is expected will allow for the investigation of the nature of dark matter, dark energy and gravity. Efforts to derive $\Lambda$ from quantum field theory face the 'cosmological constant problem', with theoretical predictions differing from observations by more than 50 orders of magnitude \cite{Martin_2012}. Moreover, from the thorough investigation of the huge amount of new high-quality data, a number of so-called tensions in cosmology have emerged, as measurements of $\Lambda$CDM parameters using different observations have yielded incompatible results, see \cite{Di_Valentino_2021, desá2022empiricalinvestigationcosmologicaltensions}. This has motivated the development of alternative scenarios, in particular modified gravity (MG) models that depart from General Relativity (GR). Such models can account for the observed accelerated expansion without requiring a cosmological constant \cite{ishak2019modifiedgravitydarkenergy}. Tests and validation of GR to date have been largely confined to small scales \cite{Jain2013ASTROPHYSICALUNIVERSE, Koyama2016CosmologicalGravity}, where in MG screening mechanisms suppress deviations from standard gravity. On cosmological scales, however, such mechanisms become ineffective, allowing potential departures from GR to manifest. Models that describe cosmic acceleration through locally screened MG can be investigated through cosmological observables at large scales (\cite{Hou:2023kfp}). Current galaxy surveys aim to probe these largest accessible scales in order to disentangle the signatures of modified gravity from those of dark energy.

The new generation of photometric galaxy surveys, such as Euclid \citep{Euclid:2024yrr} and LSST \citep{2019ApJ...873..111I}, will collect a wealth of unprecedented high-quality data. These surveys will map the 3D distribution of galaxies over increasing cosmological volumes, which are crucial for understanding both the physics of the early and late stages of the Universe evolution. Thus, these galaxy catalogues will provide new and more powerful tests of the standard model. In light of this, there is a need for reliable modelling methods to take full advantage of these developments with the goal of providing better constraints of the  $\Lambda$CDM model and possible deviations from it \citep{Abbott_2019}. In particular, we have reached a level of maturity in the data quality and theoretical tools that allows us for the first time to accurately test the assumption that the universe is described by GR on its largest scales.

One of the main avenues to test cosmological models is through the study of Large-Scale Structure (LSS) in the Universe. LSS refers to the gravitationally bound galaxies and dark matter haloes that develop into clusters and filaments as the Universe evolves. Observations of LSS involve measurements of either baryonic matter or of gravitational lensing as tracers of dark matter. Photometric galaxy clustering (GC) is a powerful probe into the dynamics of the Universe but requires modelling of the galaxy bias \citep{Kaiser1984}. The galaxy bias describes the relation between the observed galaxy distribution and the underlying matter distribution, predicted by theoretical models. To address potential degeneracies between the galaxy bias and the cosmological model parameters, additional observational probes, such as weak gravitational lensing (WL), are needed \citep{Hu2004, Bernstein2009, Joachimi2010, Nicola2016}. WL provides a direct, bias-independent measurement of the matter distribution. Additionally, galaxy-galaxy lensing (GGL), which measures the correlation between the positions of lens galaxies and the distorted shapes of background galaxies, offers a complementary handle on the nature of galaxy bias and the dependence on cosmology. This combination of correlation functions is known as 3x2pt.

The use of 3x2pt analysis has been established as a powerful tool for probing LSS in recent surveys \citep{DESY3, Heymans:2020gsg}. The 3x2pt analysis is particularly sensitive to the dynamics of LSS in the late Universe, whereas observations of the cosmic microwave background (CMB) \cite{PlanckCollaboration2018PlanckParameters} provide complementary information from the early universe. The combination of these two sets of observations, in the context of stage-IV surveys, will yield unprecendented constraints on the cosmological model.

Traditionally, tests of gravity have spanned two well-studied regimes: the Solar System scales, validated through parametrised post-Newtonian (PPN) methods \cite{Will1993}, and the largest available cosmological scales, assessed using the quasi-static approximation in perturbation theory \cite{Battye_2012, Gleyzes_2014}. However, the non-linear regime of cosmological structure formation, lying between these two regimes, remains poorly tested. Modelling this regime is computationally demanding, even in $\Lambda$CDM, and the lack of robust MG modelling techniques, beyond specific cases like $f(R)$ gravity, limits our ability to unlock the full power of cosmological observables to probe gravity.

Numerous studies in the literature employ Fisher forecasts for parametrized MG models to assess the ability of stage-IV surveys to constrain gravity \cite{casas2017linearnonlinearmodifiedgravity, Aparicio, CASAS2023101151}. These studies typically parametrise gravity across all cosmological scales using the post-Friedmann formalism \cite{Thomas_2020}, which involves a rescaling of the Poisson equation in general relativity (quantified by the parameter $\mu$) and a rescaling of the ratio of gravitational potentials in general relativity (quantified by the parameter $\eta$). These particular studies assume specific functional forms for $\mu(z)$ and $\eta(z)$ derived from established models or phenomenology. A more recent study \cite{srinivasan2024cosmologicalgravityscalesiv} has adopted a model-agnostic approach, evaluating the capacity of stage-IV surveys using 3x2pt to constrain gravity reaching non-linear scales.

In this study, however, we aim to quantify the bias in the cosmological parameters of $\Lambda$CDM caused by assuming GR when analysing galaxy mocks in a universe governed by  $f(R)$ gravity. We do this by using a set of high-fidelity full-sky simulated galaxy catalogues based on an underlying $f(R)$ Hu $\&$ Sawicki modified gravity theory. We use the two twin galaxy mocks presented in \cite{Tutusaus:2025ial} that are based on the halo catalogues by \cite{Arnold2018TheDistributions}. The twin galaxy mocks have been employed and validated in \cite{Viglione:2025tlh}. The galaxy mocks have been realised by populating the halo catalogues using a combination of extended HOD and SHAM modelling, achieving a sophisticated and highly realistic galaxy catalogue for our 3x2pt analysis. The pair of simulations used are an $f(R)$ gravity mock and a GR gravity mock with identical fiducial cosmology, initial conditions, and matched calibration against observations at low redshift. The $f(R)$ mock is characterised by the value of $f_{R0}$, which denotes the present-day background value of the scalar degree of freedom $f_R \equiv df/dR$ (with $R$ being the Ricci Scalar), which controls the strength of deviations from GR and the range of the fifth force. The galaxy mock was generated with $|f_{R0}| = 10^{-5}$, henceforth referred to as F5. Our approach is based in two steps:

\begin{itemize}

\item First we use the GR simulation as a reference case to test the validity of our methodology, to determine the scale cuts in a conservative approach and to assess possible biases in the recovered cosmology due to e.g, projection effects. 
\item Second, we repeat the exercise but replacing the GR mock data vectors with those from the $f(R)$ simulation, and quantify the cosmological parameter biases when analysing the data assuming the universe is truly described by GR.

\end{itemize}

The paper is organised as follows: in Sect. \ref{sec:Data} we describe the simulations from which we have selected our photometric galaxy samples. In Sect. \ref{sec:framework} we delineate the pipeline of inference, including the methods to measure and predict the data vectors. In Sect. \ref{sec:results} we present the comparison of the data vectors and the results of the inference analysis. In Sect. \ref{sec:discussion} we discuss our results in the context of related works. Finally, in Sect. \ref{sec:conclusion} we present our conclusions.

\section{Data}  
\label{sec:Data} 

\subsection{Halo catalogues}\label{sec:halo_catalogues}

This study utilises data from the GR and \(f(R)\) simulations presented in \cite{Arnold2018TheDistributions}. The \(f(R)\) simulation was conducted using the cosmological simulation code \textsc{mg-gadget3} \citep{Puchwein:2013lza}, a modification of \textsc{p-gadget3}, which is itself an improved version of \textsc{p-gadget2} \citep{Springel:2005mi}, that supports collision-less simulations under the Hu-Sawicki $f(R)$-gravity model \citep{Hu:2007nk}. Two collision-less cosmological simulations were generated at two different resolutions for each of the two models, the F5 model or GR. This work focuses on the higher-resolution simulations containing \(2048^3\) particles within a 768 Mpc/\(h\) box, yielding a particle mass resolution of \(M_{\rm part} = 3.6 \times 10^9 M_\odot/h\). This resolution is similar to that of the Flagship galaxy mock \citep{Euclid:2024few}, a state-of-the-art cosmological simulation for the Euclid survey, which uses a particle mass of \(M_{\rm part} = 10^9 M_\odot/h\). Both of our simulations assume the same fiducial cosmology, taken from \cite{Ade2016Planck2015Parameters}, characterised by \(\Omega_{\text{m}} = 0.3089\), \(\Omega_\Lambda = 0.6911\), \(\Omega_{\text{b}} = 0.0486\), \(h = 0.6774\), \(\sigma_8 = 0.8159\), and \(n_{\text{s}} = 0.9667\). The growth rate of the F5 simulation is defined in such a way that $\sigma_8$, which characterises clustering in the late Universe, is equal in both simulations. Effectively, the only difference in the fiducial cosmologies between the twin mocks is the $f_{R0}$ parameter.

\textsc{mg-gadget}3 employs an iterative Newton-Raphson method with multigrid acceleration on an adaptive mesh refinement (AMR) grid to solve for the scalar degree of freedom in modified gravity:

\begin{equation}
G_{\mu\nu} + f_R R_{\mu\nu} - \left( \frac{f}{2} - \Box f_R \right) g_{\mu\nu} - \nabla_\mu \nabla_\nu f_R = 8\pi G_N T_{\mu\nu}, 
\label{Eequn}
\end{equation}

where $R_{\mu\nu}$ is the Ricci tensor, the Einstein tensor is $G_{\mu\nu} = R_{\mu\nu} - \frac{1}{2}g_{\mu\nu}R$, with $g_{\mu\nu}$ being the metric tensor, $G_N$ is the gravitational constant and $T_{\mu\nu}$ is the stress-energy tensor. Instead of solving for $f_R$ directly, it calculates $u = \log(f_R / f_{R0})$ to prevent unphysical negative $f_R$ values, following the approach in \citealt{Oyaizu2008NonlinearMethodology}. The resulting $f_R$ is then used to compute an effective mass density that includes $f(R)$ effects and the chameleon mechanism \citep{Khoury:2003rn}:

\begin{equation}
    \delta \rho_{\rm eff} = \frac{1}{3} \delta \rho - \frac{1}{24 \pi G} \delta R,
\end{equation}

where $\rho$ is the perturbation of the matter-energy density and $\rho_{\rm eff}$ is the effective density perturbation. This effective density is added to the real mass density to determine the total gravitational acceleration using the Tree-PM Poisson solver in \textsc{p-gadget3}.

The simulations also produce 2D light-cone outputs comprising 400 \textsc{healpix} all-sky maps \citep{GorskiHEALPix:SPHERE}, spanning redshifts from \(z = 80\) to \(z = 0\). These evenly spaced maps, with a resolution of 805,306,368 pixels, are generated using the "Onion Universe" method \citep{Fosalba2008TheShells}. In this method, the simulation box is tiled to cover the volume up to a given redshift \(z_i\), and dark-matter particles in spherical shells centered at \(z_i\) are projected onto \textsc{healpix} maps. From these projected particle density maps, convergence maps for various redshift bins are created within the Born approximation. Using simple transformations in harmonic space, all-sky lensing maps for the deflection angle and cosmic shear can be derived \citep{Hu_2000}.

On the other hand, a 3D halo catalogue is generated as the lightcone simulation evolves. Haloes are identified using the Friends-of-Friends (FOF) algorithm in \textsc{p-gadget3} and refined with a shrinking-sphere method. The catalogue records properties such as mass, velocity, centre of mass, and tensor of inertia. Additional outputs include time slices and halo catalogues produced with the \textsc{subfind} algorithm \citep{SpringelPopulating0}.  

\subsection{Galaxy catalogues}  

In order to be able to perform a realistic cosmological analysis using the 3x2pt data-vectors, high-fidelity galaxy mocks are required. With this aim, we employ the full-sky lightcone halo catalogues obtained from the N-body simulations with galaxies that closely mimic real observations. In this section, we describe the automated calibration of the galaxy assignment pipeline described in \cite{Tutusaus:2025ial}, which was implemented through the SciPIC pipeline (\cite{CarreteroCosmoHubPlatform}) to generate the twin galaxy mocks.

First, the galaxies are added to the haloes using a combination of the Halo Occupation Distribution (HOD) model and the Sub-Halo Abundance Matching (SHAM) model. The HOD model used is characterised by free parameters that determine the minimum halo mass to host a central galaxy, $M_{\text{min}}$, the halo mass for which it contains one satellite galaxy on average, $M_1$, and the slope of the satellite mean occupation, $\alpha_{\text{h}}$. The mean number of central galaxies assigned to each halo of mass $M_{\text{h}}$ is given by:

\begin{equation}
    \langle N_{\text{cen}}\rangle = 1 \, \text{ , if }M_{\text{h}} > M_{\text{min}},
\end{equation}

whereas the number of satellite galaxies is assigned to the haloes using a Poisson distribution of mean:

\begin{equation}
    \langle N_{\text{sat}}\rangle = \left(\frac{M_{\text{h}}}{M_1} \right)^{\alpha_{\text{h}}} \, \text{ , if }M_{\text{h}} > M_{\text{min}}.
\end{equation}

The galaxy mocks have used a version of the HOD model wherein the dependence between $M_1$ and $M_{\text{min}}$ is given by a set of 7 free parameters following the work of \cite{Carretero}. Using the SHAM technique, the relation between the mass of the halo $M_\text{h}$ and the luminosity of the galaxy $L_{\text{gal}}$ is established. A scatter in the luminosity assigned to the central galaxies is applied in order to better reproduce clustering observations from surveys. The luminosity of the satellite galaxies is assigned as a function of the luminosity of the central galaxy with three additional free parameters. 

After establishing the mass-luminosity relation, the positions and velocities of the galaxies are assigned. Central galaxies are positioned at the centre of the halo and satellite galaxies' positions are assigned based on the spherical Navarro-Frenk-White (NFW) profile. A factor is introduced to reduce the distance of satellite galaxies to the centre of the halo to improve the agreement with clustering observations; more details in \cite{Tutusaus:2025ial}. Galaxy colours are assigned by dividing both the central and satellite galaxies into three populations of red, green and blue galaxies. A total of five luminosity bins are used resulting in 10 free parameters describing the fraction of galaxies in each population.

A total of 23 free parameters are used to generate the galaxy mocks from the halo catalogue. Those parameters are then calibrated to reproduce the measurements of the Sloan Digital Sky Survey (SDSS), \citep{SDSS:2002vjc}. In particular, the luminosity function and the projected correlation function $w_\text{p}$ were chosen as the observables to be used for calibration. The calibration was performed by generating a galaxy mock for each set of points in the 23-parameter space and measuring $w_\text{p}$ in it. Each realisation of the mocks is intrinsically stochastic because of the way that galaxies are assigned to the mocks, adding complexity to the calibration. The differential evolution algorithm was used to efficiently reach the best-fit parameters without requiring an excessive amount of mock realisations while minimising the effects of intrinsic variation. The calibration was performed at low redshift ($z=0.1$) and produced two galaxy mocks that reproduce the spatial distribution and clustering properties of the observed Universe, one for each model of gravity. The photometric redshifts (photo-$z$s) for the galaxy mocks are generated by adding a Gaussian dispersion to the true redshift with $\sigma = 0.05(1+z)$ to mimic the worsened photo-$z$ estimation at higher redshift. This value corresponds for example to the Euclid requirement~\citep{Euclid:2024yrr}.

\section{Framework and pipeline of analysis}\label{sec:framework}

To quantify the bias resulting from analysing an F5 Universe with a GR modelling, we need to perform an inference analysis using the 3x2pt data vectors from the F5 and GR mocks. The data vectors measured in the GR mock were used to define the scale cuts in order to exclude poorly modelled scales. The analytic data vectors to perform the two inference analysis were predicted assuming GR in the modelling and the covariance matrix was also generated analytically. Finally, we employ a Bayesian inference pipeline to estimate the posterior distribution of the cosmological parameters, and define the metrics necessary to quantify the bias in the recovered cosmology. In this section, we present an overview of the pipeline of analysis and the statistics that were utilised to measure the bias.

\subsection{Tomographic bin selection}

\begin{figure*}
\centering
\includegraphics[width=15cm]{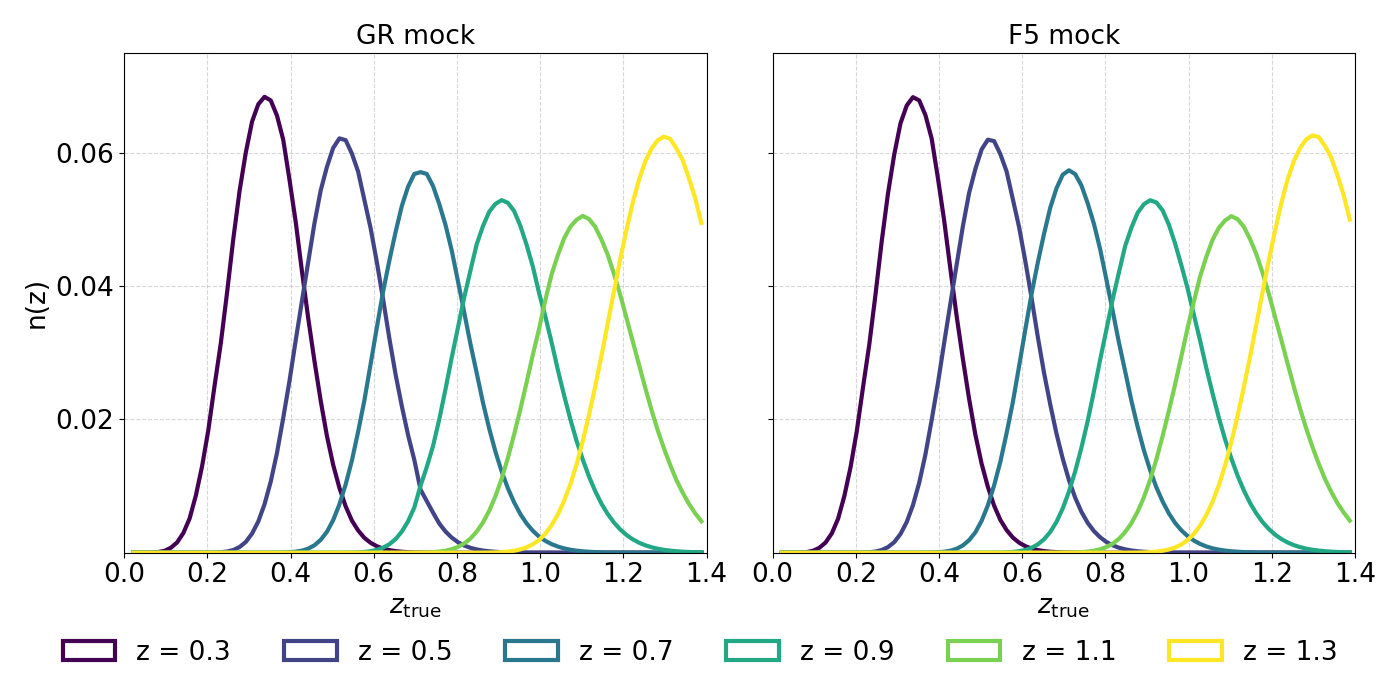}
\caption{Normalised true redshift distribution, $n(z)$, for the 6 tomographic bins. The left plot shows the distribution for the GR mock while the right plot shows the distribution for the F5 mock.}
\label{Fig:n(z)GRMG}
\end{figure*}

The 3x2pt analysis requires the classification of our galaxies into a series of tomographic redshift bins. We use the same selection criteria for the redshift bins in both catalogues. We define the bins by classifying the galaxies based on their photo-$z$s and then deriving the corresponding distribution in the so-called true redshift, which is the actual exact redshift of the galaxies in the simulation. Six redshift bins were defined, centred at redshift: 0.3, 0.5, 0.7, 0.9, 1.1 and 1.3. Each bin has a width of $\Delta z=0.2$. This choice is similar to what has been used in stage-IV forecasts like \cite{Euclid:2021osj}. We impose a cut on absolute magnitude in $0.1-r-$band of $<-18$, that is, in the absolute magnitude in the $r-$band (centred at 650 nm) for a given object if it were at a redshift of $z=0.1$. This corresponds roughly to an apparent magnitude cut of the $r-$band at $<25.5$, which effectively removes only the very faintest galaxies in the catalogue. We include all galaxies across the full sky (i.e, over a solid angle of $4\pi$ radians). In Fig. \ref{Fig:n(z)GRMG}, we present the $n(z)$ of the true redshifts of the 6 redshift bins taken for both catalogues. Note that the catalogues extend only to the limiting redshift of the simulated lightcone, $z=1.4$, which explains the abrupt truncation in the distribution. In Table \ref{tab:density3x2pt} we summarise the number of galaxies and densities for each tomographic bin considered in the analysis.  

\begin{table*}
\centering
\begin{tabular}{lcc cc}
\toprule
 & \multicolumn{2}{c}{GR} & \multicolumn{2}{c}{F5} \\ 
\cmidrule(lr){2-3}\cmidrule(lr){4-5}
Photo-$z$ bin & Nº galaxies  & Ang. density (arcmin$^{-2}$) 
              & Nº galaxies  & Ang. density (arcmin$^{-2}$) \\ 
\midrule
$0.3 \pm 0.1$ & $1.5\times 10^8$  & $0.49$ & $1.5\times 10^8$ &  $0.49$ \\  
$0.5 \pm 0.1$ & $3.0\times 10^8$  & $1.0$  & $3.1\times 10^8$ &  $1.0$  \\  
$0.7 \pm 0.1$ & $4.4\times 10^8$  & $1.5$  & $4.7\times 10^8$ &  $1.6$  \\  
$0.9 \pm 0.1$ & $5.6\times 10^8$  & $1.9$  & $5.9\times 10^8$ &  $1.9$  \\  
$1.1 \pm 0.1$ & $6.3\times 10^8$  & $2.1$  & $6.8\times 10^8$ &  $2.2$  \\  
$1.3 \pm 0.1$ & $5.3\times 10^8$  & $1.8$  & $5.8\times 10^8$ &  $1.8$  \\  
\bottomrule
\end{tabular}
\caption{Galaxy counts and angular densities in each photo-$z$ bin for the GR and F5 catalogues. The total angular density is 8.8 gal/arcmin$^{-2}$ and 8.9 gal/arcmin$^{-2}$, respectively, which is a number density that is intermediate between typical stage-III and stage-IV number densities.}
\label{tab:density3x2pt}
\end{table*}

\subsection{Measured data vectors}
\label{sec:Data Vectors}

We aim to perform an analysis of the 3x2pt angular power spectrum by calculating auto-power (A-P) and cross-power (C-P) spectra across different bins. For GC, we compute only the 6 A-P within the same bin, as most of the information is contained in the A-P. For WL, we calculate 6 A-Ps and 15 C-Ps across all combinations of bin pairs, resulting in 21 data vectors. In GGL, the distinction between lenses (density) and sources (shear) breaks the symmetry of C-Ps, leading to 30 C-Ps and six A-Ps, producing 36 data vectors. Together, this yields a total of 63 data vectors. Of these, several will be removed for the final analysis (more details in Sect. \ref{sec:scale-cuts}).

To compute the angular power spectra for a given probe $X$, (\(C_\ell^\text{X}\)), we use the publicly available code \textit{PolSpice}\footnote{\url{https://www2.iap.fr/users/hivon/software/PolSpice/}}, which requires data in the form of a HEALPix map. These maps are created by converting the right ascension (RA) and declination (DEC) coordinates into pixel indices using the \texttt{healpy} module with \texttt{nside=1024}. The galaxy count within each pixel is then used to determine the corresponding number density value. We construct a list containing the number density values for each pixel, indexed by their respective pixel numbers. As mentioned above (see Sect. \ref{sec:halo_catalogues}) we can build the convergence maps from the dark matter counts maps using the Born approximation, and other lensing observables maps, including the shear, are obtained using simple relations in spherical harmonic space. Correlating these different maps in harmonic space one can compute the GC, WL, and GGL data vectors. In more detail:

\begin{itemize}
    \item Density map: Using the number of galaxies per pixel, we can generate the density fluctuation per pixel, $\delta_{\text{pix}}$:

\begin{equation}
    \delta_{\text{pix}} = \frac{n_{\text{pix}} - \bar{n}_{\text{pix}}}{\bar{n}_{\text{pix}}},
\end{equation}

where $n_{\text{pix}}$ is the number of galaxies in that given pixel and $\bar{n}_{\text{pix}}$ is the mean number of galaxies per pixel over all the unmasked pixels.

\item Shear map: We need to generate two sets of maps corresponding to the real ($\gamma_1$) and complex ($\gamma_2$) components of the shear spin-2 field.  The values on each pixel represent the averaged shear component of all the galaxies inside that pixel. 
\end{itemize}

Since the catalogues cover the entire sky, no mask is required to exclude specific regions. However, for shear maps, which derive information from the galaxy distribution, it is necessary to handle pixels without galaxies, as these are effectively unobserved. Empty pixels correspond to unobserved regions and are masked on a per-bin basis.

The \(C^{\text{X}}_\ell\) values computed by \textit{PolSpice} are typically quite noisy, requiring binning to effectively smooth the data vectors. We achieve this by averaging the \(C_\ell\) values within each $\ell$ bin, using 10 logarithmically spaced bins between \(\ell = 30\) and \(\ell = 1000\). Before binning, we subtract the shot noise contribution, calculated as:

\[
N_{\text{shot}} = \frac{4 \pi f_{\text{sky}}}{n_{\text{gal}}} \frac{1}{\text{pixelwindow}(\ell)^2},
\]

where \(n_{\text{gal}}\) is the galaxy number density, and \(f_{\text{sky}} = 1\), given the full-sky coverage. The pixel window correction (which depends on the healpix map resolution) is included to ensure consistency with the corrections applied to the \(C_\ell\) values.

\subsection{Predicted data vectors}

The derivation of observable predictions from a cosmological model, namely from $\Lambda$CDM, involves the modelling of the evolution of matter density fluctuations in the primordial Universe. We implement a realistic 3x2pt pipeline of cosmological analysis based on the \texttt{CosmoSIS}\footnote{\url{https://cosmosis.readthedocs.io/en/latest/}} framework. \texttt{CosmoSIS}, the COSMOlogical Survey Inference System, is a cosmological parameter estimation code that allows for the implementation of custom pipelines of analysis in a flexible way. We consider a pipeline of analysis that uses the observables of 3x2pt and accounts for the galaxy bias present in the mocks.

The fluctuations in the matter density field of the Universe are related to the LSS observables of 3x2pt. The matter power spectrum quantifies the distribution of matter density fluctuations $\delta(\mathbf{x})$ across different scales. These fluctuations are defined as the excess of matter density at a given region compared with the mean matter density of the Universe:

\begin{equation}
    \delta(\mathbf{x}) = \frac{\rho(\mathbf{x}) - \bar{\rho}}{\bar{\rho}} 
\end{equation}
 
Where $\rho(\mathbf{x})$ is the matter density at position $\mathbf{x}$ and $\bar{\rho}$ is the mean density. The power spectrum $P_m(k)$ is defined as the Fourier transform of the two-point correlation function $\xi(r)$, which measures the clustering strength between two galaxies separated by a distance $r$.

\[
\left\langle \delta(\mathbf{k}) \delta^*(\mathbf{k'}) \right\rangle = (2\pi)^3 \delta_{\text{D}}(\mathbf{k} - \mathbf{k'}) P_{\text{m}}(k),
\]
 
where $\delta(\mathbf{k})$ is the Fourier transform of $\delta(\mathbf{x})$, and $\delta_D$ is the Dirac delta function. This expression defines $P_{\text{m}}(k)$ as the variance of density fluctuations per unit wavenumber. We generate the matter power spectra using the CAMB (\cite{CAMB}) implementation in \texttt{CosmoSIS}. In order to include some of the contributions from the non-linearities in the density field we include the revised Halofit corrections (\cite{halofit}) to the linear power spectrum, thus using the so-called non-linear power spectrum $P_{\text{m}}^{\text{nl}}(k)$. Although Halofit is calibrated under GR, we employ it here because our aim is precisely to assess the bias induced by analysing an F5 galaxy mock under GR assumptions. 

We divide the distribution of galaxies in our catalogues into a series of tomographic redshift bins in which to compute the observables. We predict those observables in harmonic space, with our observables in the analysis thus being the angular power spectra for GC ($C_{\ell}^{\text{gg}}$), GGL ($C_{\ell}^{\gamma \text{g}}$) and WL ($C_{\ell}^{\gamma\gamma}$). The equations for the angular power spectra of the three probes are as follows:

\begin{equation}
C_{\ell}^{\gamma\gamma, ij} = \int_0^{\chi_H} d\chi\,\frac{q^i_{\gamma}(\chi)q^j_{\gamma}(\chi)}{\chi^2}P_{\rm m}\left(\frac{\ell+1/2}{\chi},z(\chi)\right)\,,
\end{equation}

\begin{equation}
C_{\ell}^{\text{gg}, ij} = \int_0^{\chi_H} d\chi\,\frac{q^i_{\text{g}}(\chi)q^j_{\text{g}}(\chi)}{\chi^2}P_{\rm m}\left(\frac{\ell+1/2}{\chi},z(\chi)\right)\,,
\end{equation}

\begin{equation}
C_{\ell}^{\gamma \text{g},ij} = \int_0^{\chi_H} d\chi\,\frac{q^i_{\gamma}(\chi)q^j_{\text{g}}(\chi)}{\chi^2}P_{\rm m}\left(\frac{\ell+1/2}{\chi},z(\chi)\right)\,,
\end{equation}

where $\chi$ is the comoving angular diameter distance, $\chi_H$ is the comoving distance to the horizon and the indices $ij$ denote the tomographic bins. We adopt the Limber and flat-sky approximations for all spectra except for GC at $\ell<200$ where we use the full-sky spherical-Bessel integration following \cite{Fang:2019xat} (see below). The corresponding kernels are defined as:

\begin{equation}
    q^i_{\gamma}(\chi) = \frac{3 H_0^2\Omega_{{\rm m}}}{2c^2 a(\chi)}\chi \int_\chi^{\chi_h} d\chi' n^i_{\gamma} (\chi')\left(1-\frac{\chi}{\chi'} \right)\,,
\end{equation}

\begin{equation}
    q^i_{\text{g}}(\chi) = b^i(z(\chi))\, n^i_\text{g}(z(\chi)) \frac{dz}{d\chi}\,,
\end{equation}

where $b^i(z(\chi))$ is the linear galaxy bias and $n^i_\gamma (z(\chi))$ and $n^i_\text{g} (z(\chi))$ are the redshift distributions of the galaxy samples. This set of predictions for the 3x2pt angular power spectra are then used in our pipeline of parameter inference.

Regarding the computation of the power spectra, 
the Limber approximation can yield inaccurate results on large scales (low $\ell$), and therefore we performed internal consistency tests comparing the predicted data vectors from \texttt{CosmoSIS} and from the independent code Core Cosmological Library (CCL) \footnote{\url{https://ccl.readthedocs.io/en/latest/}}(\cite{pyccl}) and found that only the GC angular power spectra at $\ell<200$ deviated when using Limber. This is similar to what has been found in \cite{Fang:2019xat}. Thus, we also consider the exact calculation of the angular power spectra for GC below $\ell =200$, which takes the form as follows:

\begin{equation}
C_{\ell}^{\text{gg},ij} = \frac{2}{\pi} \int_0^\infty dk\, k^2 P_{\text{m}}^{\text{nl}}(k) \Delta_{\delta \text{g}}^i (k ,\ell)\Delta_{\delta \text{g}}^j (k ,\ell)\,,
\end{equation}
with $ \Delta_{\delta \text{g}}^2 (k ,\ell)$ being:

\begin{equation}
    \Delta^i_{\delta \text{g}} (k,\ell) = \int_0^{\chi_\text{H}} d\chi \, q^i_{\text{g}}(\chi) \,j_{\ell}(k\chi )\,,
\end{equation}
where $j_{\ell}(\chi k)$ is the Bessel function of order $\ell$.

We consider only the dominant contributions to the signal for simplicity, relying on the use of the true redshift and true positions without including redshift-space distortions and magnification into the modelling.

In terms of systematic effects, we consider the galaxy bias in our predictions. The galaxy bias parameter, as an effective scaling factor for the GC and GGL angular power spectra, is expected to be degenerate with the increased clustering of matter in an F5 Universe with accelerated gravitational collapse. Our simulated galaxy catalogues do not have an intrinsic alignment implementation, therefore its modelling is not included in our analysis. In addition, we do not consider instrumental systematics like shifts in the distribution or multiplicative bias. Our modelling follows the linear galaxy bias model and thus we concentrate our analysis to those scales not significantly affected by non-linear galaxy bias phenomena. In order to accomplish that, we define a series of scale cuts to the data vectors so as not to infer a biased cosmology due to the non-modelled non-linearities. Additionally, baryonic feedback effects could potentially have an impact on our analysis. We have discussed the implementation of baryonic effects in the power spectrum in Appendix \ref{sec:appendix_baryons}.

\subsection{Scale cuts}\label{sec:scale-cuts}

Since our goal is to quantify the bias in the recovered cosmology solely caused by assuming the wrong cosmology (i.e, GR gravity for an F5 mock universe), it is crucial to minimise the impact of other possible sources of bias. With this in mind, we have implemented very conservative scale cuts in the measured data vectors. The scale cuts have been individually defined for each probe but not for each individual data vector. We chose to include scales where the criteria was met on average for all the data vectors in a given probe. The first criteria that is implemented discards data points whenever the deviation between the GR simulation and the theory is above $5\%$. In the case of GGL, this criteria is too conservative and we opt for another criteria where we discard those scales where the theory and simulations disagree by more than $3\, \sigma$. This results in substantially more conservative scale cuts than for GC and WL but does not discard all but a few data points of GGL as the first criteria did. In the case of WL, a difference in agreement on small scales appears when testing different methods to generate the non-linear power spectrum. In particular, using Halofit produces a better agreement at high multipoles (small scales) than HMCode (\cite{Mead:2020vgs}). In light of this, we opt to err on the side of caution and determine the scale cuts based on the worst-performing modelling of the non-linear power spectrum. In addition, the data point for the lowest $\ell$-bin was discarded in all data vectors due to a signficantly low signal for many of the data vectors (arguably due to sample variance). Therefore, the resulting scale cuts for the different probes are: for GC $\ell_{min}= 40, \, \ell_{max}=400$, for GGL $\ell_{min}= 40, \, \ell_{max}=250$, and for WL $\ell_{min}= 40, \, \ell_{max}=400$. 

We include in the discussion of the scale cuts the removal of the data from two tomographic bins for our final cosmological analysis. The last tomographic bin resulted in unexpected behaviour in the WL shear measurements due to the abrupt cut of the galaxy catalogues at redshift $z=1.4$, which was not included in the theoretical modelling. In line with the idea of applying conservative scale cuts that leave little room for degeneracies in the inferred bias, we also removed the first redshift bin from the analysis due to the dominance of the non-linear galaxy bias on most scales affecting GC and GGL. The inclusion of those tomographic bins in the analysis could lead to biases in the inferred cosmology of the GR mock, so we performed the final analysis with the 4 tomographic bins centred at $z=(0.5, 0.7, 0.9, 1.1)$.

\subsection{Covariance matrix estimation} 

In order to generate the covariance matrix that we consider in the analysis we use the public code \texttt{OneCovariance}\footnote{\url{https://github.com/rreischke/OneCovariance}} (\cite{OneCovariance}). We compute the Gaussian terms of the covariance with the addition of the Super Sample Covariance (SSC) (\cite{Takada:2013wfa}), which describes the inherent variance introduced by surveys of limited volume not encapsulating density fluctuation at scales larger than the survey volume. We also include non-Gaussian contributions (nG) to the covariance despite being subdominant at the scales included in the analysis. We opted to use an analytic covariance to avoid numerical instabilities that can arise using a Jackknife-estimated covariance. 

The Gaussian terms of the covariance arise from the assumption that the underlying matter density field distribution is Gaussian. With this assumption, $\ell$ modes are considered to be uncorrelated and the Gaussian covariance matrix can be calculated as:

\begin{align}
     &\textbf{Cov}_{G} (C_{\ell}^{ij}, C_{\ell'}^{kl}) = \frac{1}{(2\ell+1)\,f_{\text{sky}} \Delta \ell} \cdot \notag\\ 
    &\left[(C_{\ell}^{ik} + N_{\text{shot}}^{ik})(C_{\ell}^{jl} + N_{\text{shot}}^{jl}) + (C_{\ell}^{il} + N_{\text{shot}}^{il})(C_{\ell}^{jk} + N_{\text{shot}}^{jk})\right] \delta_{\ell \ell'},
\end{align}
where the upper indexes refer to the tomographic bins. The $\Delta \ell$ term appears because we are using multipole bins instead of computing for each individual $\ell$ and $N_{\ell}$ is the shot noise for each probe, given by:

\begin{align}
N_{\text{gg}}^{ij} &= \frac{1}{\bar{n}^i}\delta_{ij}, \\
N_{\gamma \text{g}}^{ij}  &= \frac{\sigma_e^2}{\bar{n}^i}\delta_{ij}, \\
N_{\gamma\gamma}^{ij}  &= \frac{\sigma_e^2}{\bar{n}^i \bar{n}^j},
\end{align}
where the ellipticity dispersion per component is $\sigma_e = 0.23$ in line with Stage-IV surveys (\cite{Euclid_blanchard}), $\bar{n}^i$ is the galaxy number density of the $i$-th bin and $\delta_{ij}$ is the Kronecker delta. 

The SSC contribution, arising from the low $\ell$ modes not being fully captured due to the limited volume of the survey, affects mostly the lensing and clustering probes at large scales. It introduces uncertainty at lower multipoles and is described by:

\begin{equation}
    \textbf{Cov}_{\text{SSC}} (C_{\ell}^{ij}, C_{\ell'}^{kl}) = \frac{\partial C_{\ell}^{ij}}{\partial \delta_\text{b}}\frac{\partial C_{\ell}^{kl}}{\partial \delta_\text{b}} \sigma_\text{b}\,,
\end{equation}
where $\delta_\text{b}$ refers to the background density fluctuations in the matter field at scales beyond the survey and $\sigma_\text{b}$ is the variance of that fluctuation field. The derivatives quantify the sensitivity of each angular power spectra to the background modes. 

Finally, the non-Gaussian terms of the covariance $\textbf{Cov}_{\text{nG}}$ arise from trispectrum contributions to the uncertainty. We refer to \cite{OneCovariance} for further details on the derivation of these terms. In summary, the full covariance that we use in our inference pipeline is the sum of these three contributions:

\begin{equation}
    \textbf{Cov} = \textbf{Cov}^{\text{G}} + \textbf{Cov}^{\text{SSC}} + \textbf{Cov}^{\text{nG}}\,.
\end{equation}

\begin{figure}
    \centering
    \subfloat[Predicted Gaussian correlation matrix generated with \texttt{OneCovariance}.]{\includegraphics[width=0.475\textwidth]{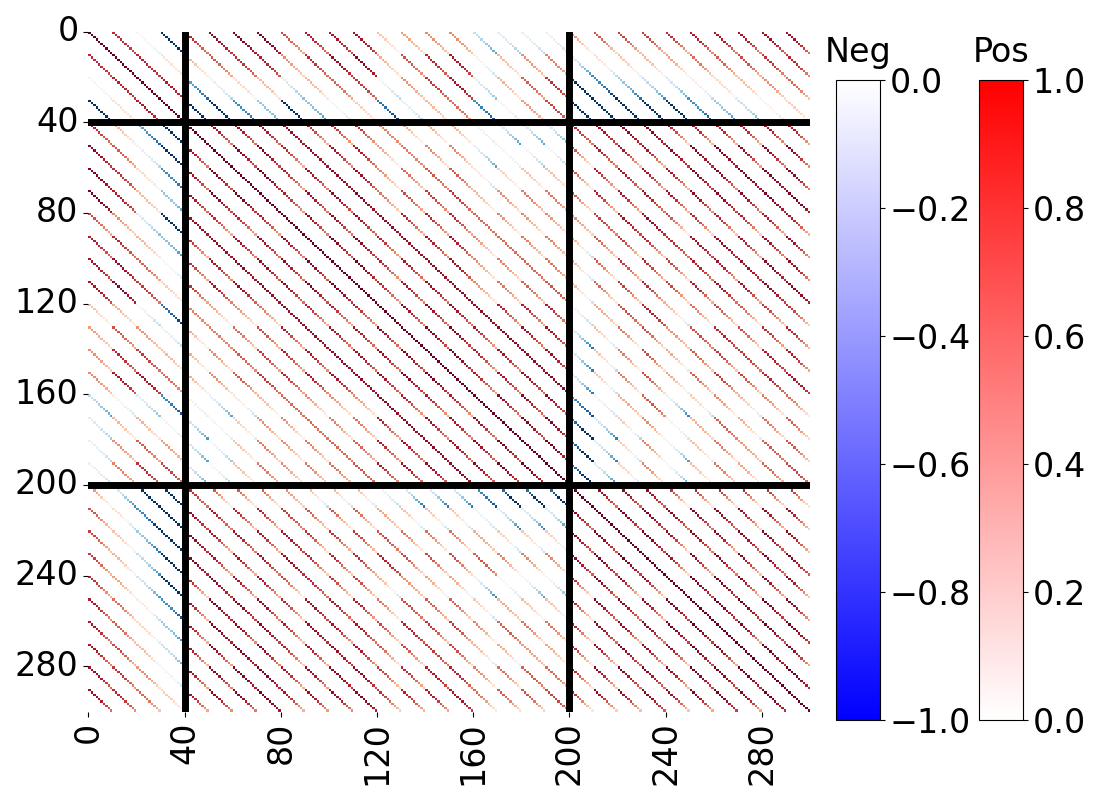}\label{fig:onecov_gaussian}}
    \hfill
    \subfloat[Predicted correlation matrix including SSC and nG generated with \texttt{OneCovariance}.]{\includegraphics[width=0.475\textwidth]{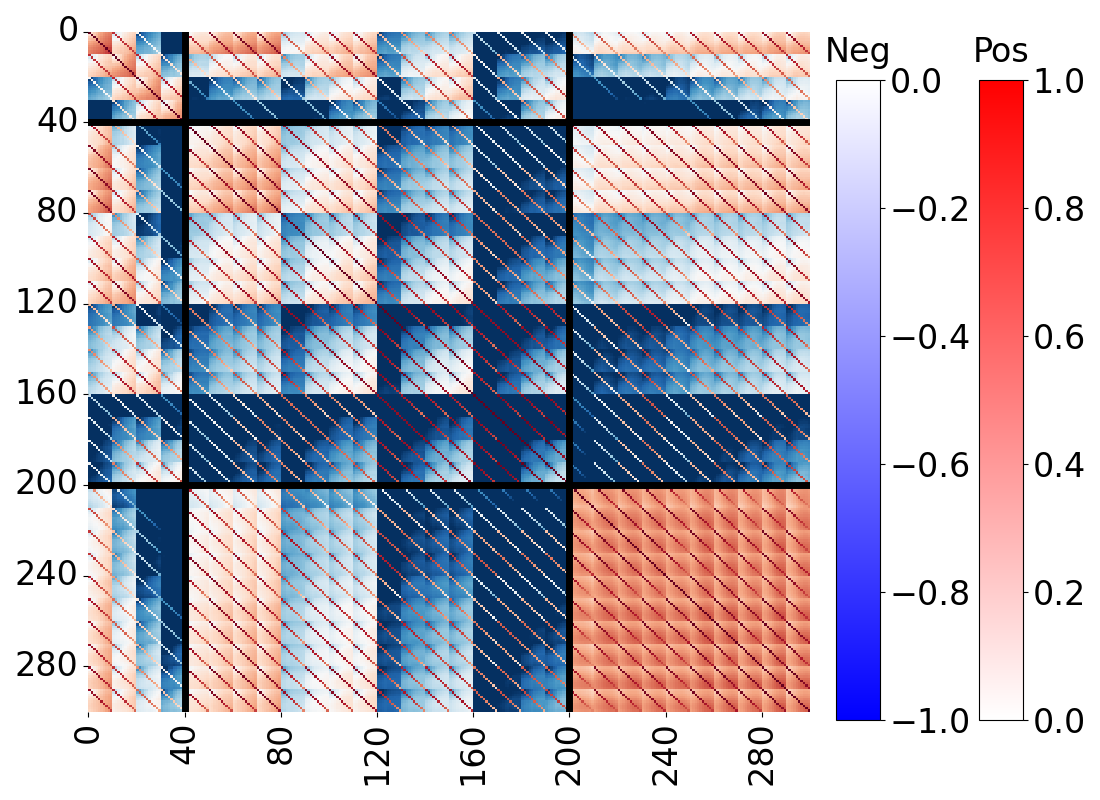}\label{fig:onecov_ssc_ng}}
    
    \caption{Analytic correlation matrices for the 3×2pt data vectors, computed with \texttt{OneCovariance}. (a) Gaussian-only terms. (b) Full covariance including Gaussian, super-sample covariance (SSC), and non-Gaussian (nG) contributions. The matrix is block-structured into GC, GGL, and WL observables, from left to right.}
    \label{fig:cov_compare}
\end{figure}

In Fig. \ref{fig:cov_compare} we can explicitly see the contribution of SSC and nG to the off-diagonal terms of the covariance. We see in Fig. \ref{fig:onecov_gaussian} how the Gaussian covariance only includes the diagonal terms, given the lack of mode mixing in that model prediction. In contrast, as seen in Fig. \ref{fig:onecov_ssc_ng}, the SSC off-diagonal terms introduce mostly positive correlations in WL and negative correlations in GGL and, to a lesser extent, GC. In addition, the nG term contributes both to positive non-diagonal and diagonal terms of the covariance, increasing variance at high $\ell$. This can be seen as the positive off-diagonal terms at the high $\ell$ regions of each submatrix. However, this contribution is subdominant at the scales we use, below $\ell =1000$.   

\subsection{Inference methods}\label{sec:inference_methods}

Our last step in the analysis is the estimate of the posterior distribution of the model parameters. Our parameter estimation framework is based on Bayesian inference, which allows for the estimation of the uncertainties on cosmological parameters. In the Bayesian formalism, the posterior probability distribution for our model parameters $\boldsymbol{\theta}$ given a set of observed data $\boldsymbol{d}$ is estimated from the likelihood $\mathcal{L}(\boldsymbol{d}|\boldsymbol{\theta})$ using Bayes' theorem:

\begin{equation}
P(\boldsymbol{\theta}|\boldsymbol{d}) = \frac{\mathcal{L}(\boldsymbol{d}|\boldsymbol{\theta}) \pi(\boldsymbol{\theta})}{\mathcal{Z}}\,,
\end{equation}
where $\pi(\boldsymbol{\theta})$ represents our prior beliefs about the parameters and $\mathcal{Z} = \int \mathcal{L}(\boldsymbol{d}|\boldsymbol{\theta}) \pi(\boldsymbol{\theta}) d\boldsymbol{\theta}$ is the Bayesian evidence.
 
Our model testing is based on nested sampling, which is an efficient method for computing the posterior distribution in high-dimensional parameter spaces. Nested sampling has been extensively used in cosmology, notably in cosmological surveys like  DES-Y3 (\cite{DESY3}), KiDS (\cite{KiDS-Y4}) or Planck (\cite{Planck2018}). A sophisticated implementation of a nested sampling algorithm is Polychord (\cite{polychord}), as implemented in the \texttt{CosmoSIS}\footnote{\url{https://cosmosis.readthedocs.io/}} framework (\cite{CosmoSIS}). Polychord is a robust nested sampling algorithm suited to sample multi-modal posteriors and high-dimensional models and it has been used in the cosmological inference pipelines for DES-Y3 data. We employ Polychord with internal parameters $n_{live}=100$, tolerance=0.1 and default settings for the rest of the configuration. The posterior distributions for different sets of parameters generated with Polychord are then used to quantify the bias in the cosmology through the use of several metrics.

\subsection{Metrics}

Accurately quantifying the bias in the recovered cosmology of the F5 galaxy mock requires consideration of the appropriate metrics. We shall use two complementary metrics to evaluate the bias: the typically used Figure of Bias (FoB) to estimate the statistical significance of the bias and the Probability to Exceed (PTE) to estimate the expected bias that could appear due to the variance present in a single galaxy mock.

The quantification of a bias in the inferred marginalised 2-dimensional posterior distributions is performed using the FoB. The marginalised 1-dimensional posteriors alone can fail to capture biases for highly correlated parameters. The FoB is defined as follows:

\begin{equation}
    \text{FoB} = \sqrt{\left(\vec{\mu}-\vec{x} \right)^T \textbf{Cov}^{\text{Post}} \left(\vec{\mu}-\vec{x} \right)}\,,
\end{equation}

\begin{equation}
    \textbf{Cov}^{\text{Post}} = \begin{bmatrix}
\sigma_{x_i,x_i} & \sigma_{x_i,x_j}  \\
\sigma_{x_j,x_i} & \sigma_{x_j,x_j}\\
\end{bmatrix}\,,
\end{equation} 
where $\vec{x}$ is the set of mean values in our roughly Gaussian posteriors, $\vec{\mu}$ is the set of fiducial values used in the simulations and $\textbf{Cov}^{Post}$ is the covariance matrix from the combined 2-dimensional posterior for a pair of variables.

The discussion of appropriate metrics needs deep consideration (see for example \cite{DES_metrics}), as metrics like the FoB combine deviations originating from projection effects, common in high-dimensional models, along with actual deviations from the theory model. We discuss these projection effects in more detail in Appendix \ref{sec:appendix_validation}. In addition to that, being limited to a single mock for our measurements creates a statistical noise such that we may recover a non-zero bias even in the absence of any systematic effects.     

In order to mitigate these problems in the interpretation of the bias in the recovered cosmology, we employ the same recipe as in \cite{Blake_2024}. We consider the PTE, which quantifies the likelihood that the best-fit cosmology will deviate from the fiducial cosmology of the mock at a given level of statistical significance and independently of projection effects. 

We start by generating a set of synthetic noiseless data vectors from our model, $\vec{d}_{\text{synth}}$, and we run our Bayesian inference pipeline for them. Any projection effects which may arise due to our covariance matrix or analysis choices (such as the scales included in the analysis or our modelling) will be present in this case as well. We choose to focus on $\Omega_{\text{m}}$ and $\sigma_8$ as our parameters of interest. The iso-likelihood contours from the 2-dimensional marginalised posterior at a given $n-\sigma$ level will define the region $R\,(n-\sigma)$ in the \{$\Omega_{\text{m}}, \sigma_8$\} parameter space.

We then evaluate how much of the posterior derived from the mock data lies within this region:

\begin{equation}
    P\,(\Omega_{\text{m}},\sigma_8|\mathbf{d}_{\text{mock}})_{R(n-\sigma)} = \int_{R(n-\sigma)} P\,(\Omega_{\text{m}},\sigma_8|\mathbf{d}_{\text{mock}})\,d\,\Omega_{\text{m}} \,d\,\sigma_8,
\end{equation}

which quantifies the fraction of the mock posterior contained within the $n\text{-}\sigma$ contour of the synthetic posterior. This measure expresses how consistent the recovered cosmology from the mock is with the expectation from the model. The PTE is then computed for a given $n\text{-}\sigma$ level as:

\begin{equation}
    \text{PTE}\,(n-\sigma) = 1 - P\,(\Omega_{\text{m}},\sigma_8|\mathbf{d}_{\text{mock}})_{R(n-\sigma)}.
\end{equation}

The interpretation of the PTE should be that it represents the likelihood that an $n-\sigma$ deviation between the best-fit cosmology and the fiducial cosmology will take place due to the inherent variability in having a limited number of mock realisations. As an illustrative example, if we computed the PTE by substituting the data vectors measured in the mock, $\vec{d}_{\text{mock}}$, for the ones used in the prediction, $\vec{d}_{\text{synth}}$, we would recover PTE values of $(0.32, 0.05, 0.003, ...)$ for $(1\sigma, 2\sigma, 3\sigma,...)$ respectively, regardless of projection effects. A PTE value close to 1 would indicate a high likelihood that a deviation in the recovered cosmology solely due to the noise of the mock could be found at a given $n-\sigma$ level and a value close to zero would indicate that a deviation at that $n-\sigma$ level is unlikely to occur as a statistical fluke. 

The PTE quantifies the expected deviation in the recovered cosmology without being affected by projection effects. If projection effects were present, as we find they are in Appendix \ref{sec:appendix_validation}, they would result in a higher deviation than expected by the PTE alone, so the PTE should be understood as a threshold deviation due only to noise due to the limited number of mock realisations. 

With the assumption that the posterior distributions are Gaussian, the FoB would be related to the $n-\sigma$ of the PTE. In practise, the posteriors are not Gaussian and the FoB does not exactly correspond to the an iso-likelihood contour for which to compute the PTE. Still, the consideration of both the FoB to compute cosmological bias and of the PTE to estimate the likelihood of a bias at a given statistical level complement each other. We use both metrics to assess the bias in the cosmology that we infer.

\section{Results}\label{sec:results}

\begin{figure*}
    \centering
    \includegraphics[width=\textwidth]{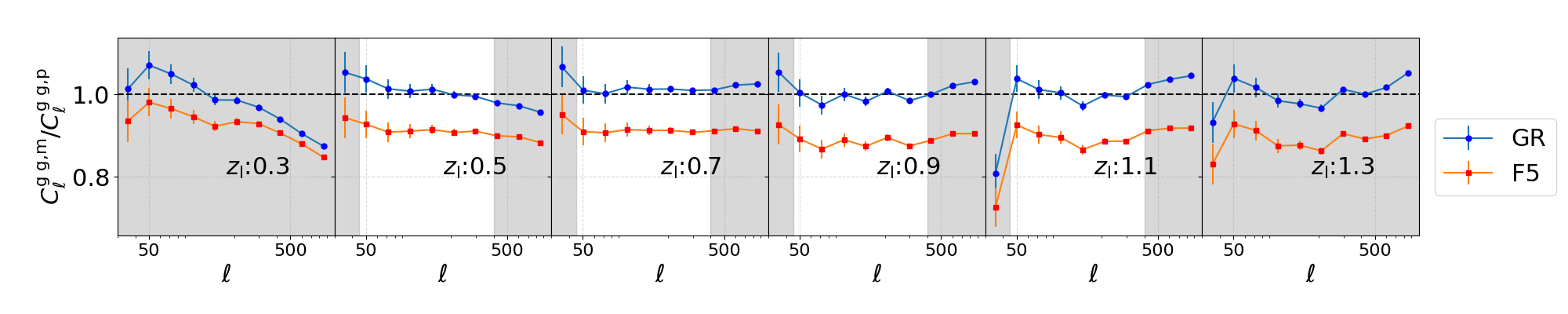}
    \caption{Ratio between the measured data vectors (superscript m) in the GR and F5 mocks and the analytic data vectors (superscript p) of GC generated with our model and the fiducial cosmology of the mock. The label $\text{z}_\text{l}$ refers to the redshift of the centre of the tomographic bin.}
    \label{fig:GC_dvs}
\end{figure*}

\begin{figure*}
    \centering
    \includegraphics[width=\textwidth]{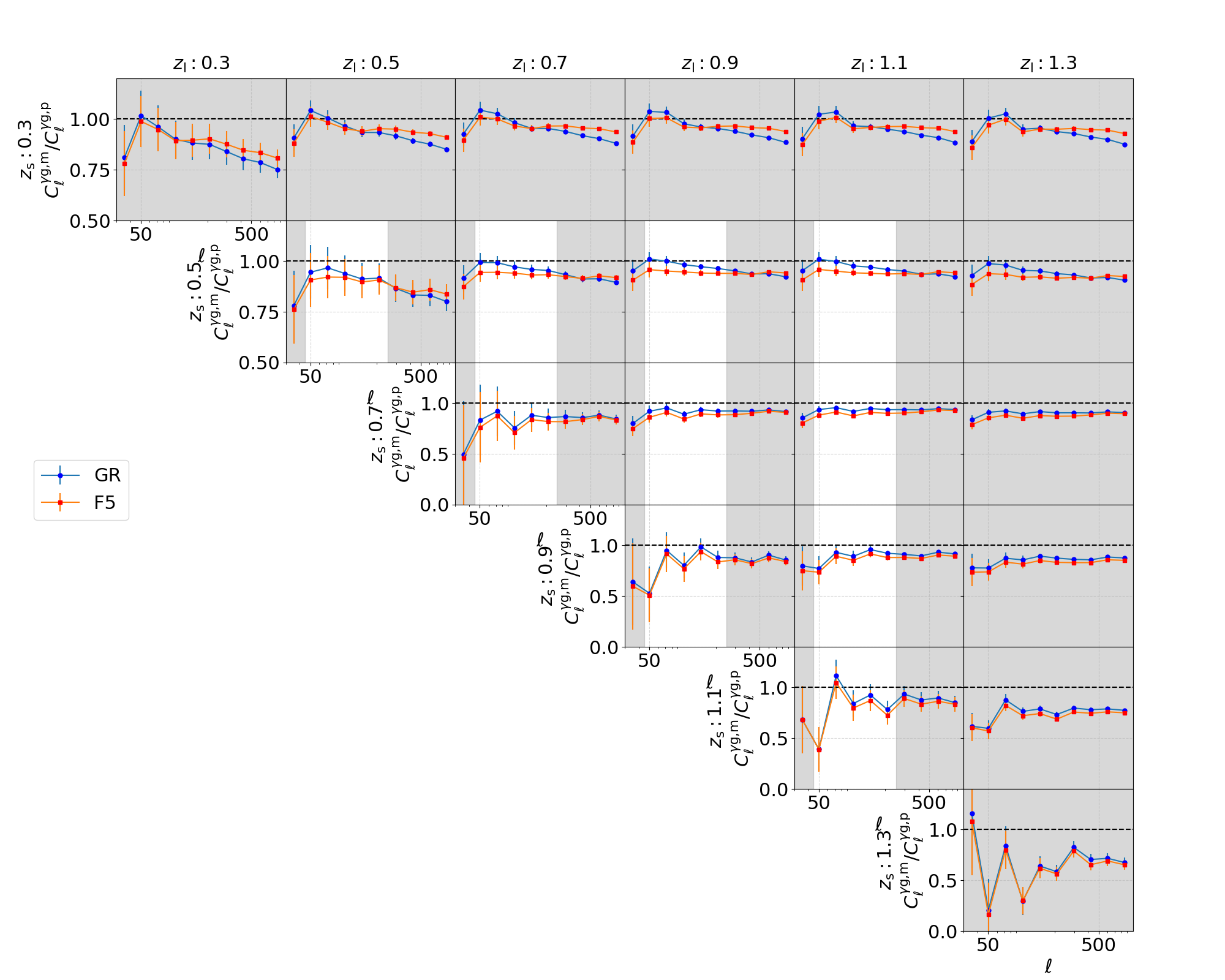}
    \caption{Ratio between the measured data vectors (superscript m) in the GR and F5 mocks and the analytic data vectors (superscript p) of GGL generated with our model and the fiducial cosmology of the mock. The labels $\text{z}_\text{s}$ and $\text{z}_\text{l}$ refer to the redshift of the centre of the tomographic bin for the source and lens galaxies, respectively.}
    \label{fig:GGL_dvs}
\end{figure*}

\begin{figure*}
    \centering
    \includegraphics[width=\textwidth]{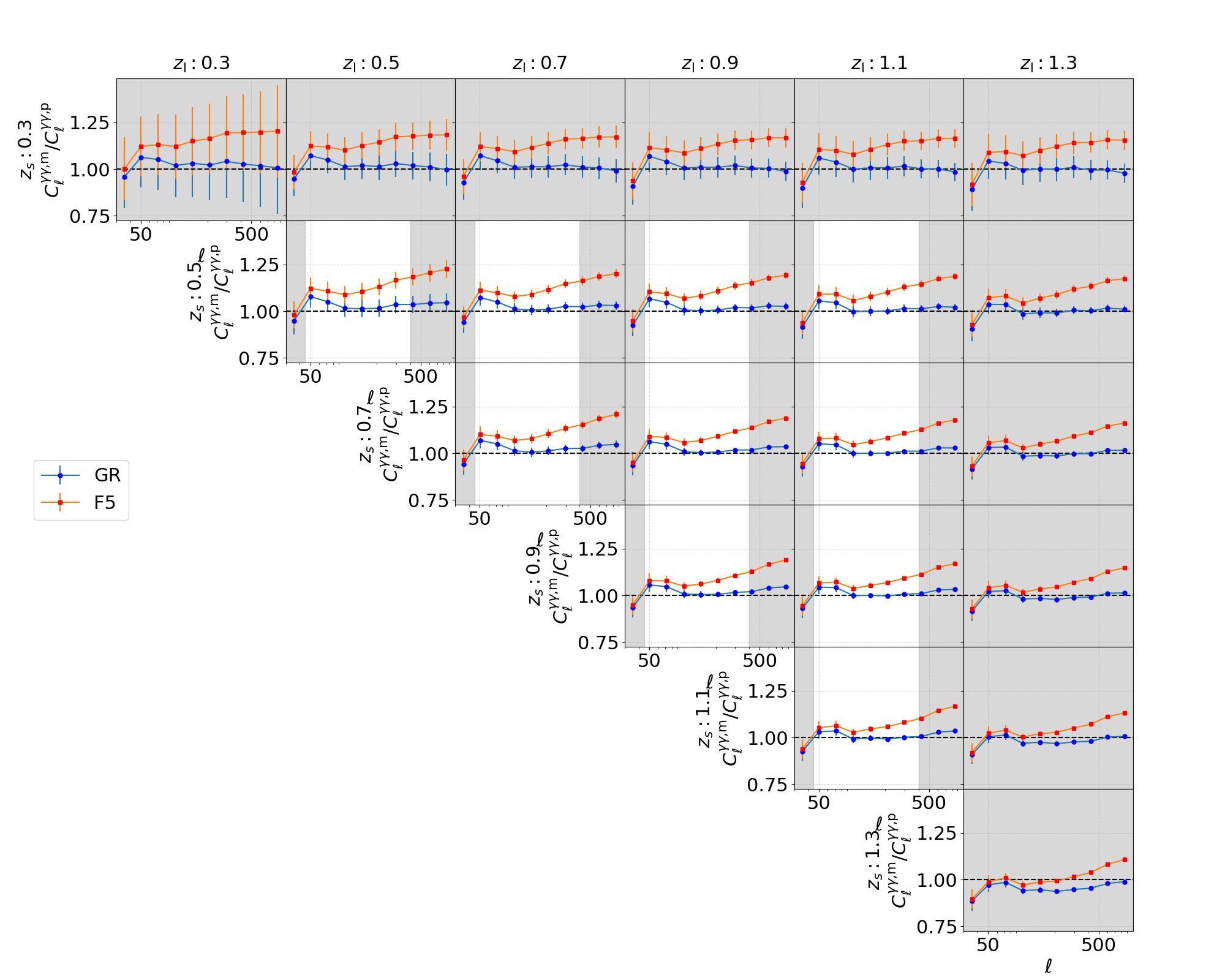}
    \caption{Ratio between the measured data vectors (superscript m) in the GR and F5 mocks and the analytic data vectors (superscript p) of WL generated with our model and the fiducial cosmology of the mock. The labels $\text{z}_\text{s}$ and $\text{z}_\text{l}$ refer to the redshift of the centre of the tomographic bin for the source and lens galaxies, respectively.}
    \label{fig:WL_dvs}
\end{figure*}

\subsection{Data vector comparison}

In this section we present the data vectors by plotting the ratio between the values for measurements in the F5 and GR mocks and the theoretical predictions from our analytical model, which assumes GR (i.e, $\Lambda$CDM). The uncertainties are given by the analytic covariance matrix.

In Fig. \ref{fig:GC_dvs}, we present the results for GC using the same value of the galaxy bias for both sets of data vectors. The GR mock exhibits a stronger clustering signal compared to F5. The difference between the GC data vectors measured in the GR and F5 mocks is approximately $10\%$ in all bins except for the $z=0.3$ bin. At small scales in the $z=0.3$ redshift bin, both data vectors are most similar, which aligns with expectations since the mocks were calibrated at low redshifts and small scales, leading to similar clustering \cite{Tutusaus:2025ial}. In the other redshift bins, the F5 mock consistently shows lower clustering. Consequently, we expect the inferred galaxy bias to be lower for the F5 mock compared to the GR mock. The scale cuts are shown in grey and contain the data points discarded in the analysis.

In Fig. \ref{fig:GGL_dvs}, we present the GGL data vectors. The ratio between measured and predicted data vectors is similar in both the GR and F5 mocks, with GR showing a slightly stronger signal at large scales, while the opposite trend is observed at small scales. In this case, it is not sufficient to change the amplitude of the (linear) galaxy bias for the two measured data vectors to match due to the differing slope, unlike for GC. 

Lastly, in Fig. \ref{fig:WL_dvs}, we present the WL data vectors. In this case, the signal is stronger for F5 compared to GR, with the difference becoming more pronounced at smaller scales, as expected from the accelerated growth of structure in this particular modified gravity model with respect to $\Lambda$CDM.

\subsection{Parameter inference}

We run the analysis using the data vectors measured from the simulations. We perform parameter inference for two different sets of cosmological parameters, namely $(\Omega_{\text{m}}, \sigma_8)$ and $(\Omega_{\text{m}},\Omega_{\text{b}}, \sigma_8, h, n_{\text{s}})$. This choice is based on the fact that the 3x2pt analysis is mainly sensitive to $\Omega_{\text{m}}$ and $\sigma_8$, with surveys of LSS using 3x2pt like DES-Y3 results focusing on those parameters \citep{DESY3}. We apply CMB-based priors in the latter case to account for the reduced constraining power of LSS probes for parameters $\Omega_{\text{b}}$, $n_{\text{s}}$ and $h$. Our only systematic effect present in the measured data vectors, the galaxy bias, is included in the sampling. In this section we compare the bias in the inferred cosmology when modelling gravity correctly assuming GR for a GR Universe or incorrectly analysing an F5 Universe under the GR assumption. 

\begin{figure*}
    \centering
    \includegraphics[width=0.90\linewidth]{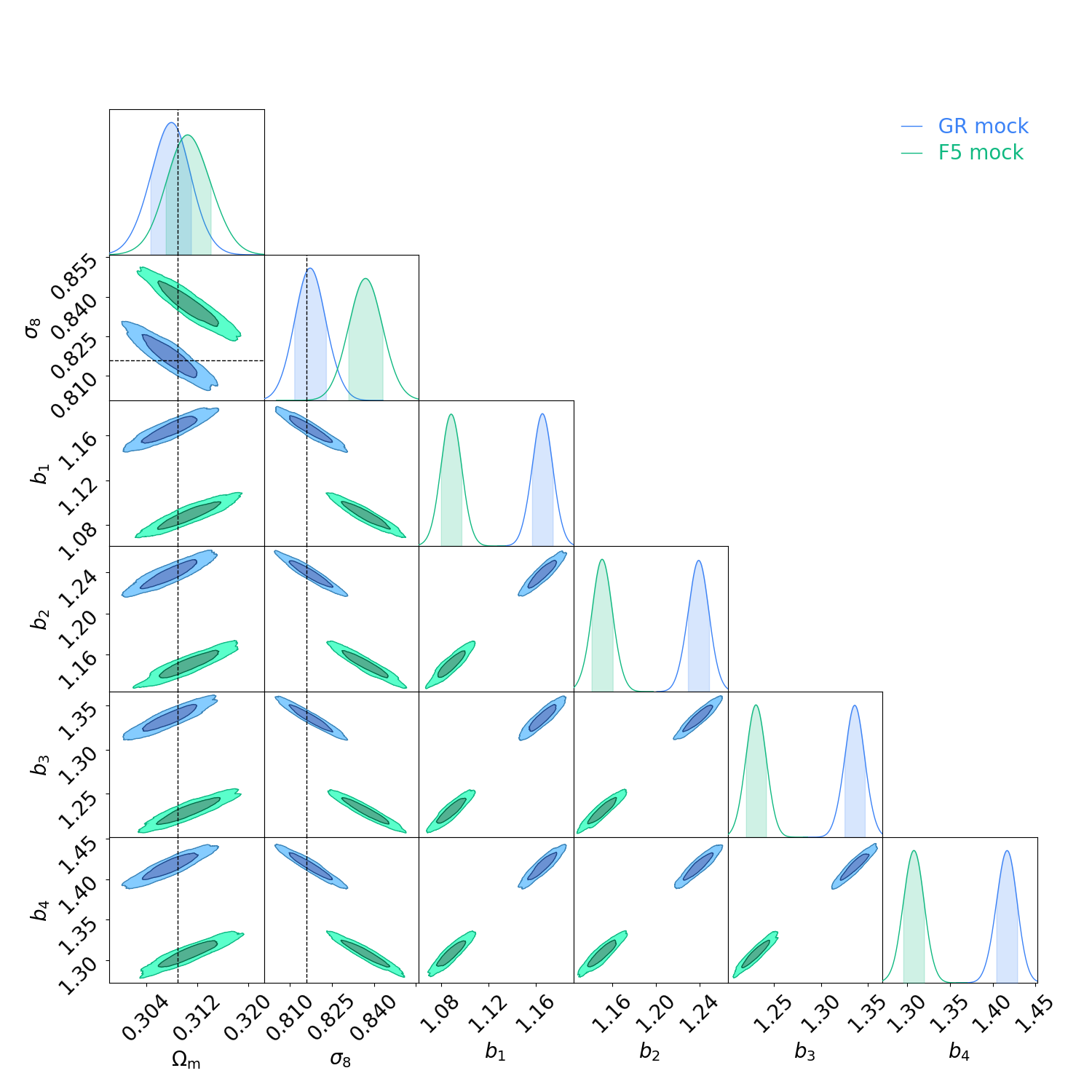}  
    \caption{Constraints on all the sampled parameters in our 3x2pt analysis for our $\Omega_{\text{m}}$ and $\sigma_8$ case. Here the galaxy bias constraints are shown explicitly. The input values of the cosmology of the simulations are shown as black lines.}
    \label{fig:2param_wgbias}
\end{figure*}

In Fig. \ref{fig:2param_wgbias} we show the constraints for $\Omega_{\text{m}}$ and $\sigma_8$ when using the GR Universe measurements. We recover the input cosmology of the simulation as the true values of the parameters fall within the $1-\sigma$ contour with a deviation of $0.4\sigma$. Looking at Table \ref{tab:pte_values}, the PTE at $1-\sigma$ and at $2-\sigma$ for this case are 0.52 and 0.16, respectively, indicating that a small deviation at that statistical level would be expected.

In contrast, when the data vectors of the F5 Universe are used, the inferred cosmology is incompatible with the input cosmology with a FoB of $\approx12\sigma$. In this case, the PTE at $5-\sigma$ of $4.3 \times 10^{-5}$ indicates that a bias of $5\sigma$ or more is very unlikely to be explained by the variance in using a single mock realisation. The increased structure formation in an F5 Universe could result in an overestimation of the matter content of the Universe, thus overestimating $\Omega_{\text{m}}$, or an overestimation of the amplitude of the density fluctuations, thus overestimating $\sigma_8$. For this reason, while marginalised 1-dimensional posteriors are compatible with the fiducial values for these two parameters, their highly correlated 2-dimensional posteriors show a very significant bias. The assumption of GR in the modelling of an F5 Universe would therefore lead to a poor characterisation of its underlying cosmology, even when utilising very conservative scale cuts. 

 The galaxy bias parameters are also expected to differ in an F5 Universe due to the accelerated clustering of matter. Both the overall dark matter distribution and the halo distribution have enhanced clustering but the screening mechanism reduces this effect for the more massive haloes. Since the galaxies are assigned based on the halo catalogue, the decreased halo bias also results in a decreased galaxy bias. This trend is evident in the panels for the galaxy bias parameters for each tomographic bin in Fig. \ref{fig:2param_wgbias}, where the inferred galaxy bias in an F5 Universe is smaller than in GR with high statistical significance. Since both the galaxy bias parameters and $\sigma_8$ act as a scaling factor for the GC and GGL data vectors, it would be reasonable to expect a bias in $\sigma_8$ to be degenerate with a shift in the galaxy bias parameters. The fact that both the galaxy bias and the cosmological parameters $\Omega_{\text{m}}$ and $\sigma_8$ are biased altogether showcases the strength of the cosmological bias, as the shift in cosmological parameters cannot be absorbed by adjusting a nuisance parameter.

\begin{figure}
    \centering
    \includegraphics[width=0.90\linewidth]{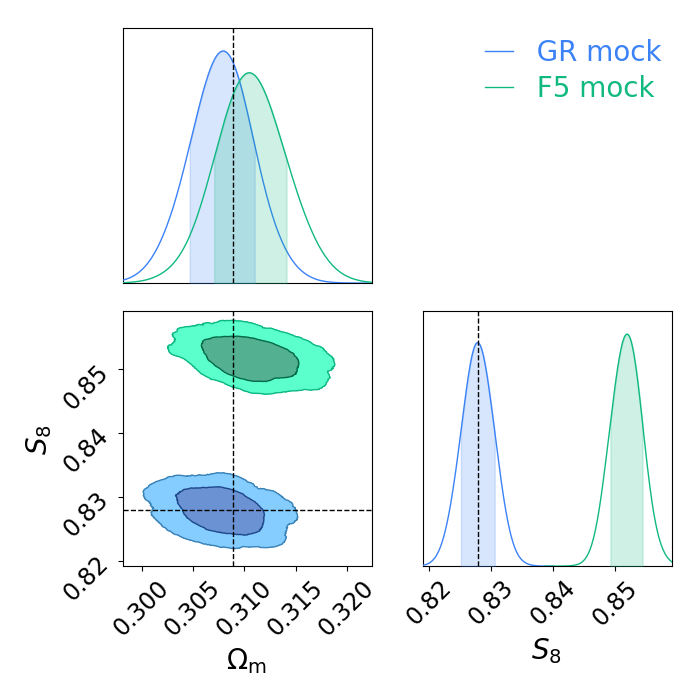}  
    \caption{Constraints on the derived parameter $S_8$ and $\Omega_{\text{m}}$. The corresponding values for the cosmology of the simulations is shown as the two black lines.}
    \label{fig:2param_wgbias_s8_omm}
\end{figure}

In our analysis we focus on the $S_8$ parameter, which is a measure of the degree of inhomogeneity or structure growth in the Universe, defined as $S_8 = \sigma_8 \sqrt{\Omega_{\text{m}}/0.3}$. The $S_8$ parameter has been specifically constructed to capture the degeneracy between $\Omega_{\text{m}}$ and $\sigma_8$ and is therefore useful in quantifying the bias that we observe for those two parameters. In Fig. \ref{fig:2param_wgbias_s8_omm} we present the constraints on $S_8$ as well as with $\Omega_{\text{m}}$. Our results indicate that while the fiducial values of parameters for the simulation are mostly recovered within the $2-\sigma$ level for the GR Universe, we find that the recovered $S_8$ in the F5 Universe is significantly biased. In particular, the deviation for the F5 mock universe is estimated to be FoB $\approx12\sigma$. 

Comparing the bias in the 2-dimensional posterior of $\Omega_{\text{m}}$ and $\sigma_8$, shown in Fig. \ref{fig:2param_wgbias}, to the bias in the 1-dimensional posterior of $S_8$, displayed in Fig. \ref{fig:2param_wgbias_s8_omm}, we can see that the biases are similar. This is because $S_8$ is particularly suited to represent the differing growth of structure between the two mocks thanks to its dependence on both $\Omega_{\text{m}}$ and $\sigma_8$ simultaneously. Measuring such a significant bias in $S_8$ emphasises the strength of the bias resulting from incorrectly assuming GR in an F5 Universe. 

While our simulation does not include baryonic effects, the question of whether baryonic effects can mask the so-called "$S_8$ tension", i.e, the apparent discrepancy in measurements of such parameter inferred from cosmic probes and low and high redshift, has been recently investigated (see e.g, \cite{Terasawa2024ExploringTB}, \cite{Amon:2022azi}). In the case of \cite{Terasawa2024ExploringTB} the inclusion of baryonic effects in emulator-derived observables was not enough to resolve the tension. The work of \cite{Amon:2022azi} considered only non-linear scales in a WL analysis but the baryonic feedback suppression of the power spectrum at those scales was not sufficient to resolve the $S_8$ tension either. In Appendix \ref{sec:appendix_baryons} we explore the impact of adding baryonic feedback to the modelling of both the GR and F5 simulation using our setup to investigate whether the cosmological biases would be masked by such effects. We find that the addition of baryonic feedback modelling, which is enclosed in an additional nuisance parameter, does not absorb or reduce the observed cosmological bias. 

Although the best constrained parameters in LSS probes are those most related to structure formation, with photometric clustering being less sensitive to the shape of the primordial power spectrum or to the Hubble parameter, it is important to understand how the bias in $\Omega_{\text{m}}$ and $\sigma_8$ changes when the rest of the parameters can vary freely in order to assess the robustness of this bias. In order to assess this, we have repeated the above analysis but sampling over a a larger set of cosmological parameters, $\Omega_{\text{m}}$, $\Omega_{\text{b}}$, $\sigma_8$, $h_0$ and $n_{\text{s}}$, which are the basic set in the $\Lambda$CDM model. 

Acknowledging that the additional set of parameters, $\Omega_{\text{b}}$, $h_0$ and $n_{\text{s}}$, are not significantly constrained by the 3x2pt photometric clustering probes, we choose to introduce priors on them based on the Planck CMB data. 
The Gaussian priors included in the \texttt{CosmoSIS} Bayesian inference pipeline are as follows: $\left(\mu_{\Omega_{\text{b}}},\sigma_{\Omega_{\text{b}}} \right) = \left(4.86\times10^{-2}, 2.2\times10^{-4}\right)$, $\left(\mu_{n_{\text{s}}},\sigma_{n_{\text{s}}} \right) = \left(9.67\times10^{-1}, 4.0\times10^{-3}\right)$ and $\left(\mu_{h_0},\sigma_{h_0} \right) = \left(6.77\times10^{-1}, 5.0\times10^{-3}\right)$.

\begin{figure*}
    \centering
    \includegraphics[width=0.9\linewidth]{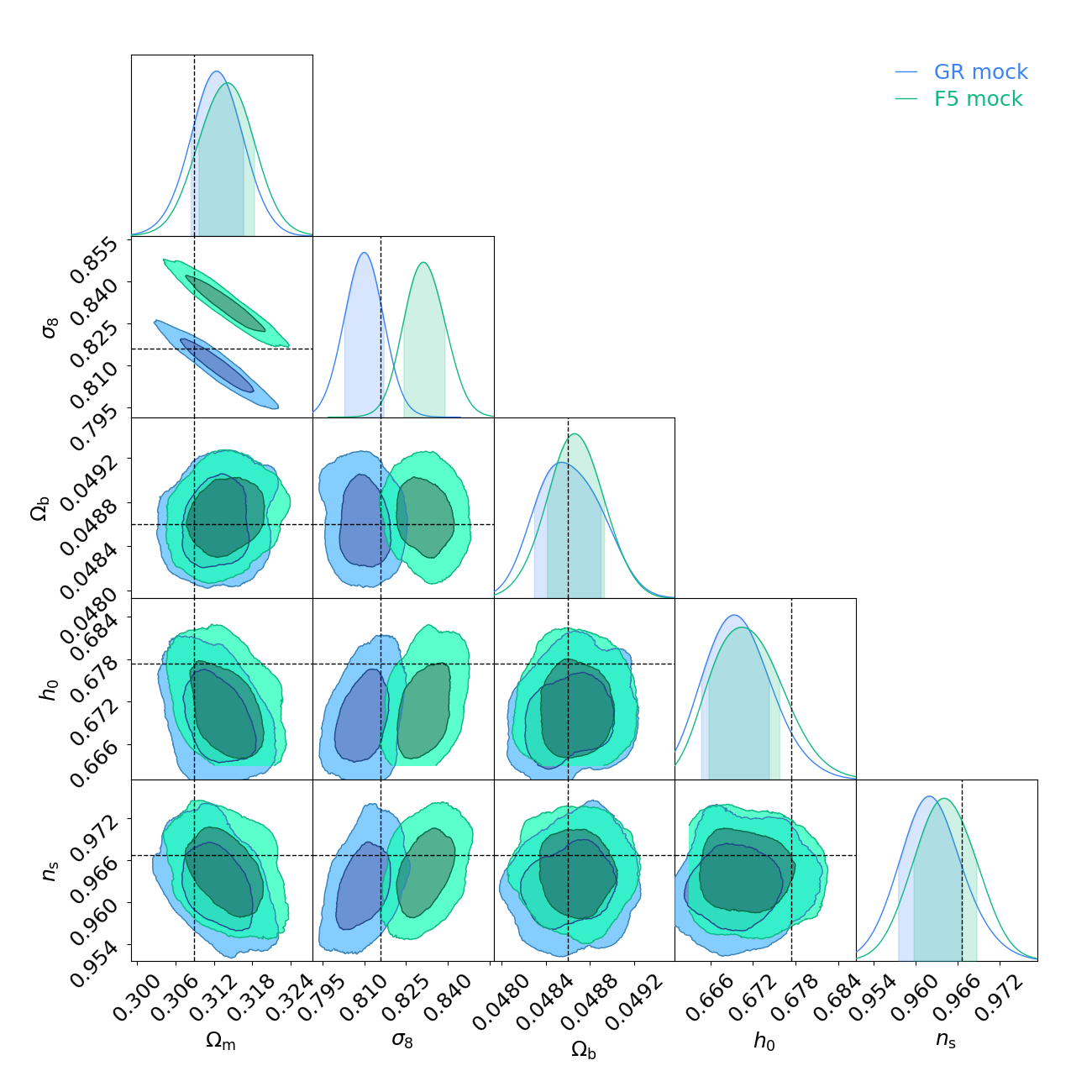}  
    \caption{Constraints for the basic set of $\Lambda$CDM cosmological parameters, assuming CMB-based (Planck) priors for $h_0$, $n_{\text{s}}$ and $\Omega_{\text{b}}$.}
    \label{fig:2param_wgbias_all}
\end{figure*}

In Fig. \ref{fig:2param_wgbias_all} we see the constraints on the basic set of $\Lambda$CDM cosmological parameters recovered from our pipeline of inference. In the GR case, we find an FoB of $1.1\sigma$, which would \textit{a priori} suggest an incompatibility with the fiducial cosmology of the mock. However, contextualising that FoB value with the corresponding values of the PTE from Table \ref{tab:pte_values} of $1-\sigma$ and $2-\sigma$, 0.39 and 0.094, respectively, we see that a deviation between $1\sigma$ and $2\sigma$ could be reasonably attributed to the noise in the mock realisation alone. Considering that the projection effects will further bias the cosmology, more so in this case with more parameters in our modelling, we find that such deviation in the GR case is not statistically significant. Therefore we conclude that the cosmology of the GR mock is well recovered within expectations. In contrast, for the case of the F5 galaxy mock, the FoB is of $\approx 12\sigma$, which is far beyond any expected deviations due to noise or projection effects, and thus a significant bias that might be potentially detected with Stage IV experiments.

\begin{table*}
\centering
\caption{PTE values for different confidence levels in the marginalised \{$\Omega_{\text{m}} , \sigma_8$\} posterior.}
\label{tab:pte_values}
\begin{tabular}{lcccc}
\hline
PTE for the marginalised \{$\Omega_{\text{m}} , \sigma_8$\} posterior & $1\sigma$ & $2\sigma$ & $3\sigma$ & $5\sigma$ \\ 
\hline
Fixed \{$\Omega_{\text{b}},n_{\text{s}}, h_0$\} & 0.52 & 0.16 & 0.024 & 4.3$\times 10^{-5}$ \\  
Free \{$\Omega_{\text{b}},n_{\text{s}}, h_0$\} & 0.39 & 0.094 & 0.011 & 2.1$\times 10^{-5}$ \\  
\hline
\end{tabular}
\end{table*}

\section{Discussion}\label{sec:discussion}

The combined analysis of current observational data on the largest cosmological scales suggests that the simplest $\Lambda$CDM model might be in crisis. In particular, the so-called $S_8$ parameter that quantifies the growth of structure inferred from different probes does not seem to converge to a statistically consistent value.
In this context, our 3x2pt analysis of the $S_8$ bias induced by a wrong assumption of the underlying gravity theory may be of particular interest in the possible resolution of such cosomlogical tension. LSS probes yields to lower values for $S_8$ (KiDS: $S_8=0.758 \pm 0.02$,\cite{KiDS-Y4}, DES-Y3: $S_8=0.776 \pm 0.017$, \cite{DESY3}, HSC: $S_8=0.785 \pm 0.03$, \cite{HSC:2018mrq}) while the Planck measurements prefer a higher value ($S_8=0.834 \pm 0.016$, \cite{Planck2018}). It is worth mentioning that the $S_8$ tension is no longer present in the results of the KiDS DR5 analysis (\cite{Wright:2025xka}), where they combined the KiDS DR5 data with that of other surveys. While the bias that we found for $S_8$ is in the opposite direction, our results suggest that modelling the Universe with the wrong gravity model can have a very substantial impact on the inferred value of $S_8$. This sensitivity of $S_8$ to gravity could be further explored in future work within the context of the $S_8$ tension. 

On the other hand, a more complete analysis of the base cosmological parameters of the $\Lambda$CDM model, such as $\Omega_{\text{b}}$, $h_0$ and $n_{\text{s}}$, suggests that a combined photometric clustering in a stage IV experiment, might not be constraining enough to detect such cosmological biases. We note that this conclusion is drawn from our particular analysis choices i.e, when using 4 tomographic redshift bins, conservative scale cuts, and CMB priors. However, recent work by \cite{pca_mg} suggests that the scale cuts chosen for the analysis of modified gravity data could be optimized through a PCA methodology. In their approach they argue they significantly improve constraints while reducing biases in the recovered cosmology. The implementation of those optimized scale cuts into our pipeline could potentially ease the bias we detect, but it is unlikely that it can bring the F5 mock universe measurements in agreement with the underlying GR model,  given the very high statistical significance of the deviations. 

The work presented in \cite{Harnois-Deraps:2022bie} also is highly relevant to the results that we present. They used WL simulations constructed with ray-tracing over an N-body simulation with $f(R)$ gravity. When performing a Bayesian inference analysis assuming GR in the modelling of the $f(R)$ simulation they recovered a catastrophic bias of $\approx20\,\sigma$ for the $S_8$ parameter. Their analysis was limited to the WL probe instead of 3x2pt and they used simulated WL maps instead of fully realised galaxy mocks. Their work is complementary to ours, with the bias they obtained in the recovered cosmology reinforcing the main result of our work: the parameters of MG are highly degenerate with clustering parameters and the assumption of an incorrect model of gravity will result in catastrophic biases.

Alternatively, \cite{euclid_forecast_fofR} used synthetic data vectors derived from $f(R)$ emulators, and found no cosmological biases when assuming a $\Lambda$CDM model. The contrast with other works poses the question of whether synthetic data vectors derived from $f(R)$ emulators can capture the cosmological bias that is obtained when using realistic galaxy simulations. Notably, baryonic effects were also included in their modelling, which could potentially mask the effects of $f(R)$ gravity. In contrast, we find that including baryonic modelling in our analysis does not reduce the cosmological bias in the recovered cosmology of F5 (more details in Appendix \ref{sec:appendix_baryons}). We attribute this difference to our use of realistic galaxy mocks with which we have derived our 3x2pt data vectors, arguably capturing differential features of the GR and $f(R)$ gravity observational probes that are not fully characterised with the synthetic data vectors.

\section{Conclusions}\label{sec:conclusion}

The impact of modified gravity models on the 3x2pt probes remains largely unexplored. In our work, we have presented the first comparative analysis using a high-fidelity galaxy mock with full-sky coverage based on an $f(R)$ modified gravity model N-body simulation. Our analysis focuses on the so-called 3x2pt probes, involving GC, WL and GGL. Although a speed up in the growth of structure formation is expected in such modified gravity models, it is not evident whether this accelerated growth can significantly bias the inferred cosmology with 3x2pt analysis in the context of Stage-IV experiments. In this paper we have shown that this is indeed possible, making use of state-of-the-art galaxy mocks that combine the volume and mass resolution needed to faithfully reproduce the galaxy properties expected from leading current and future astronomical surveys such as Euclid or LSST. The galaxy mock generated with F5 gravity, Hu-Sawicki gravity with $|f_{R0}| =10^{-5}$, is accompanied by a twin mock generated assuming GR instead, but otherwise with the same initial conditions and cosmological parameters. This GR mock is used to define the conservative scale-cuts that are implemented and it is also used to measure the statistical significance of the bias in the inferred cosmology of the F5 mock. The use of a sophisticated F5 galaxy mock to measure the 3x2pt data vectors for an inference analysis is novel and provides a realistic estimate of the cosmological bias that can result from assuming GR in the modelling of a MG universe.

Assuming GR as our reference model of gravity when analysing an F5 mock universe results in statistically significant biases in the parameters most sensible to LSS in the late Universe, $\Omega_{\text{m}}$ and $\sigma_8$. The inferred bias for the marginalised 2-dimensional posterior of $\Omega_{\text{m}}$ and $\sigma_8$ is of $12\sigma$, and a similar bias arises in the inferred $S_8$ parameter. Remarkably, the $S_8$ parameter, assuming GR in an otherwise F5 Universe, would result in a value that is much larger than expected, which should be easily detected by Stage-IV surveys. Moreover, we have shown that this bias is present even when including potentially degenerate systematic effects like the galaxy bias or baryonic feedback.

The high-fidelity galaxy mocks used have allowed, for the first time, to develop an end-to-end analysis for Stage-IV-like surveys in terms of the realistic modelling of galaxy properties (including galaxy density, redshift depth and sky coverage, as well as a calibration against observational data at low redshift). In summary, our results show that a Bayesian inference analysis using the 3x2pt observables is susceptible to generate a strong cosmological bias if GR were to be assumed when modelling an $f(R)$ Universe. With the wealth of information that is expected from the new generation of surveys, MG theories should be carefully considered in future analyses as cosmological tensions may be related to the gravity theories that are assumed in the modelling.

\begin{acknowledgements}
The authors acknowledge support form the Spanish Ministerio de Ciencia, Innovaci\'on y Universidades, projects PID2019-11317GB, PID2022-141079NB, PID2022-138896NB; the European Research Executive Agency HORIZON-MSCA-2021-SE-01 Research and Innovation programme under the Marie Skłodowska-Curie grant agreement number 101086388 (LACEGAL) and the programme Unidad de Excelencia Mar\'{\i}a de Maeztu, project CEX2020-001058-M. IT has been supported by the Ramon y Cajal fellowship (RYC2023-045531-I) funded by the State Research Agency of the Spanish Ministerio de Ciencia, Innovaci\'on y Universidades, MICIU/AEI/10.13039/501100011033/, and Social European Funds plus (FSE+).
This work has made use of CosmoHub, developed by PIC (maintained by IFAE and CIEMAT) in collaboration with ICE-CSIC. It received funding from the Spanish government (grant EQC2021-007479-P funded by MCIN/AEI/10.13039/501100011033), the EU NextGeneration/PRTR (PRTR-C17.I1), and the Generalitat de Catalunya. Ayuda PRE2020-094899  de la ayuda financiada por MCIN/AEI/10.13039/501100011033 y por FSE invierte en tu futuro.
\end{acknowledgements}

\bibliographystyle{mnras}
\bibliography{biblio} 
\clearpage
\appendix

\section{Baryonic feedback}\label{sec:appendix_baryons}

While LSS is dominated by gravity at large scales, structure formation is influenced by baryonic physics at non-linear scales. In particular, the most dominant contribution from baryonic physics comes from Active Galactic Nuclei (AGN) outflows of gas, reducing star formation and suppressing structure formation at small scales. This results in a suppression of the clustering and weak lensing signal at those small scales. In contrast, F5 gravity accelerates structure formation, increasing clustering at small scales. Our mock simulations lack baryonic physics and so the expectation is that the addition of baryonic effects to the modelling should not affect the inferred cosmology for the GR mock. Still, we investigated whether the addition of baryonic feedback modelling could mask the bias in the recovered cosmology in F5. 

The dominant effect of baryonic physics, AGN-driven outflows, is parametrised in the HMCode model of the non-linear power spectrum by the temperature of the gas expelled from the haloes, $T_{AGN}$. Higher values of this temperature would imply higher rates of ejection of gas from the haloes and thus higher suppression at small scales. 

\begin{figure*}
    \centering
    \includegraphics[width=0.90\linewidth]{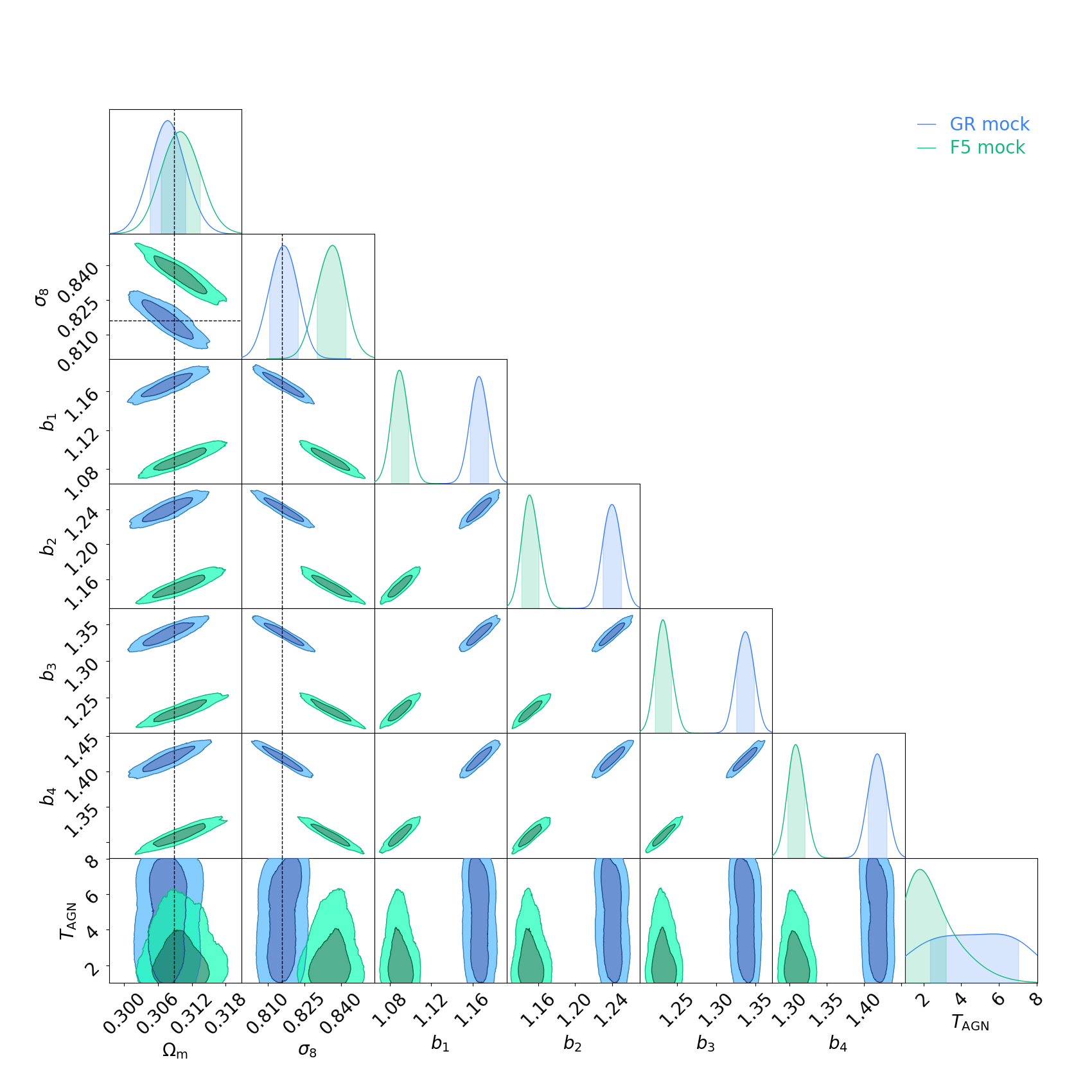}  
    \caption{Constraints on all the sampled parameters in our 3x2pt analysis sampling $\Omega_{\text{m}}$ and $\sigma_8$ together with the galaxy biases and with the addition of baryonic effects to the modelling.}
    \label{fig:2param_wgbias_baryons}
\end{figure*}

In Fig. \ref{fig:2param_wgbias_baryons} we show the contour plots analogous to Fig. \ref{fig:2param_wgbias} with the addition of baryonic feedback in the modelling. The inclusion of an additional degree of freedom results in a better fit between the inferred cosmology and the fiducial cosmology of the mock for the GR case. However, the recovered cosmology for the F5 mock remains highly biased with a similar deviation of $\approx 10\sigma$ in the \{$\Omega_{\text{m}}, \sigma_8$\} marginalised posterior. This indicates that the addition of baryonic effects to the modelling will not mask or absorb the bias in the inferred \{$\Omega_{\text{m}}, \sigma_8$\} for F5.

\begin{table*}
\centering
\caption{PTE values for different confidence levels in the marginalised \{$\Omega_{\text{m}} , \sigma_8$\} posterior modelling for baryonic feedback effects.}
\label{tab:pte_values_baryons}
\begin{tabular}{lcccc}
\hline\hline
PTE for the marginalised \{$\Omega_{\text{m}} , \sigma_8$\} posterior & $1\sigma$ & $2\sigma$ & $3\sigma$ & $5\sigma$ \\ 
\hline
Fixed \{$\Omega_{\text{b}},n_{\text{s}}, h_0$\} & 0.44 & 0.11 & 0.012 & 3.3$\times 10^{-6}$ \\  
Free \{$\Omega_{\text{b}},n_{\text{s}}, h_0$\} & 0.48 & 0.12 & 0.011 & 6.7$\times 10^{-7}$ \\  
\hline
\end{tabular}
\end{table*}

\begin{figure*}
    \centering
    \includegraphics[width=0.90\linewidth]{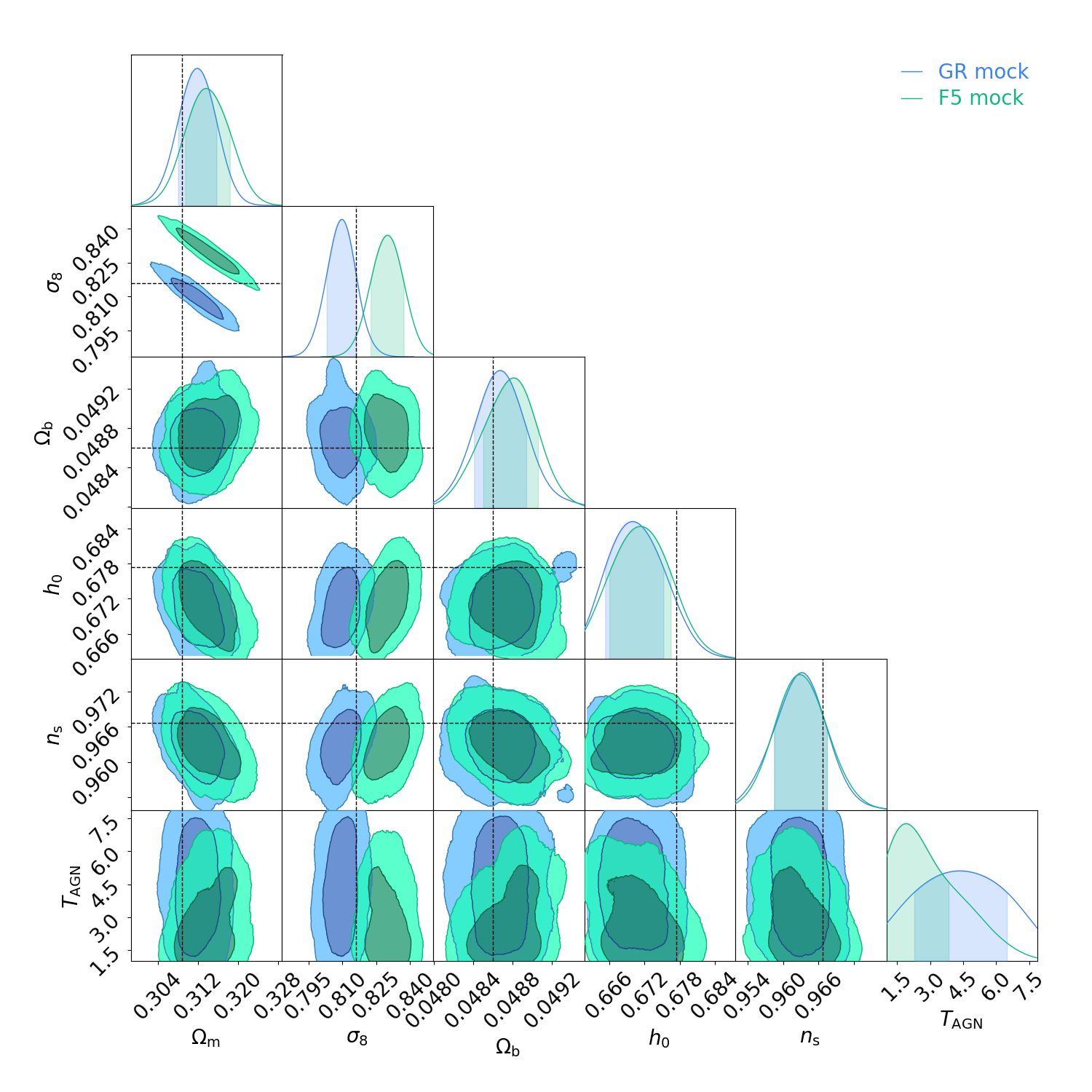}  
    \caption{Constraints on all the sampled parameters in our 3x2pt analysis for our $\Omega_{\text{m}}$ and $\sigma_8$ case with the addition of baryonic effects to the modelling.}
    \label{fig:2param_wgbias_baryons_all}
\end{figure*}

In Fig. \ref{fig:2param_wgbias_baryons_all} we show the case where we include \{$\Omega_{\text{b}}, h_0, n_{\text{s}}$\} to the sampling. Similarly, the fiducial cosmology of the mock is better recovered in the GR case compared to the analogous case without baryonic modelling, as expected after the addition of an unconstrained degree of freedom. The bias in the recovered cosmology for the F5 mock persists with a similar bias at $\approx 11\sigma$. In both cases, the FoB in the GR case is consistent with the high values of the PTE at $1\sigma$, $2\sigma$ and $3\sigma$, as seen in table \ref{tab:pte_values_baryons}. These results suggest that the addition of baryonic modelling should not, \textit{a priori}, mask a bias in the recovered cosmology resulting from modelling an F5 Universe with GR.

\section{Pipeline validation and projection effects}\label{sec:appendix_validation}

Having a large number of parameters in our model increases the possibility of projection effects biasing our inferred cosmology. The potential for projection effects generating a bias in the cosmology are not captured by the PTE. Here, we present a discussion of projection effects. 

We produce the marginalised 2-dimensional contours for a set of synthetic noiseless 3x2pt data vectors generated by the same model that is used in our \texttt{CosmoSIS}-based pipeline of analysis. By running our pipeline on these predicted data vectors, we can assess the impact of our choices regarding the covariance matrix, scale cuts and priors on the posterior distributions. 

\begin{figure*}
    \centering
    \includegraphics[width=0.90\linewidth]{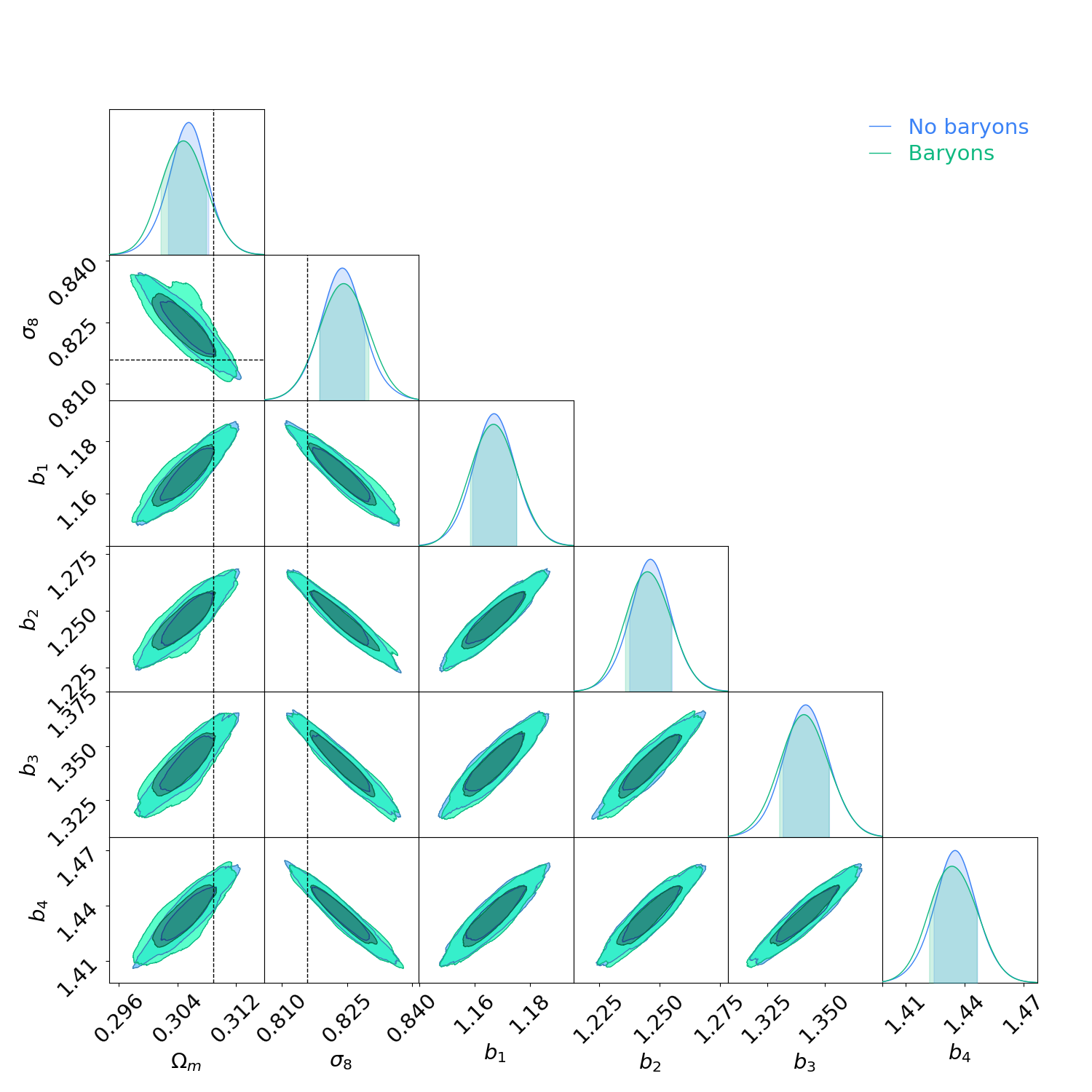}  
    \caption{Marginalised constraints generated by running the cosmological parameter inference pipeline on synthetic noiseless data vectors, sampling \{$\Omega_{\text{m}},\sigma_8$\}. The green contours correspond to the case where baryonic feedback is included in the modelling while for the blue contours there is no baryon modelling in the pipeline of analysis. Projection effects are the only source of bias in the recovered cosmology.}
    \label{fig:2param_wgbias_synth}
\end{figure*}

In Fig. \ref{fig:2param_wgbias_synth} we present the posteriors for these synthetic and noiseless data vectors modelled with or without including baryonic effects. The sampled parameters are $\Omega_{\text{m}},\, \sigma_8,$ the galaxy bias parameters and additionally $T_{AGN}$ for the case where baryonic feedback is included in the model. While the fiducial cosmology of the data vectors is recovered, a certain amount of projection effects can be seen for most of the parameters, where the fiducial values differ from the best-fit despite remaining within $1\sigma$ or $2\sigma$ at most. Although these effects are of little statistical significance by themselves, the combination of projection-based biases and mock realisation variance-based biases can result in statistically significant deviations between the fiducial and recovered cosmologies for the GR mock-measured data vectors.

\begin{figure*}
    \centering
    \includegraphics[width=0.90\linewidth]{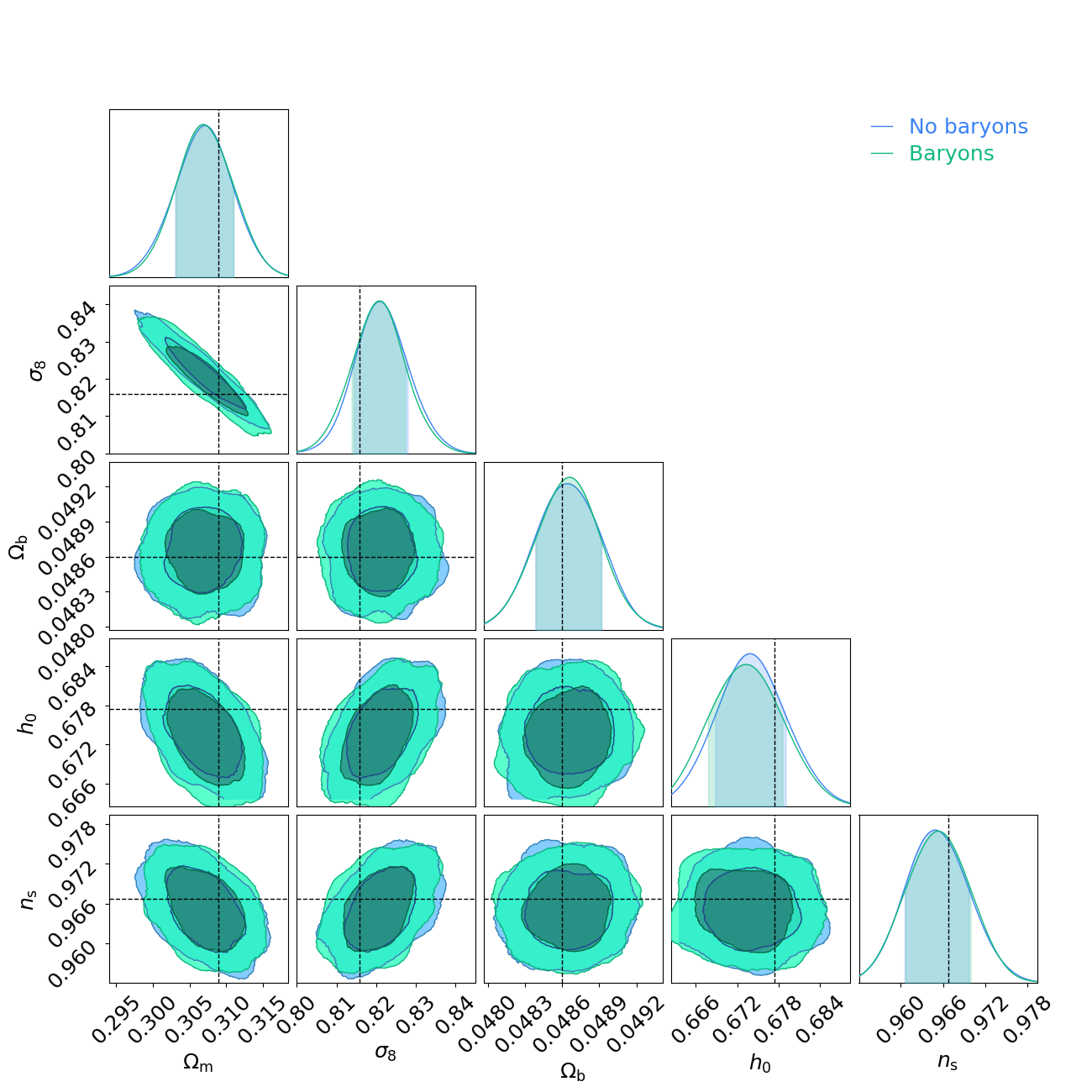}  
    \caption{Marginalised constraints generated by running the cosmological parameter inference pipeline on synthetic noiseless data vectors, sampling \{$\Omega_{\text{m}},\sigma_8,\Omega_{\text{b}},n_{\text{s}},h_0$\}. The green contours correspond to the case where baryonic feedback is included in the modelling while for the blue contours there is no baryon modelling in the pipeline of analysis. Projection effects are the only source of bias in the recovered cosmology.}
    \label{fig:2param_wgbias_synth_all}
\end{figure*}

In Fig. \ref{fig:2param_wgbias_synth_all} we show the result of including \{$\Omega_{\text{b}},n_{\text{s}},h_0$\} in the sampling, along with the corresponding Planck-derived priors. Since the extra parameters are heavily constrained by the priors the marginalised contours are not degraded. However, some projection effects can be observed as shifts between the fiducial and the best-fit values for parameters like $\Omega_{\text{m}}$, $\sigma_8$ or $h_0$.

\end{document}